\pdfoutput=1
\documentclass[fleqn,usenatbib]{mnras}

\usepackage{newtxtext,newtxmath}
\usepackage[T1]{fontenc}
\usepackage{ae,aecompl}
\usepackage{graphicx}
\usepackage{amsmath}
\usepackage{amssymb}
\usepackage{subfigure}
\usepackage{hyperref}
\usepackage{times}
\usepackage[dvipsnames]{xcolor}

\title[The Sheet of Giants]{The Sheet of Giants: Unusual Properties of the Milky Way's Immediate Neighbourhood}

\author[M. K. Neuzil et al.]{
Maria K. Neuzil,$^1$\thanks{E-mail: maria.neuzil@gmail.com}
Philip Mansfield$^{2,3}$
and Andrey V. Kravtsov$^{2,3,4}$
\\
$^{1}$Department of Physics, The University of Saint Thomas, Saint Paul, MN 55105, USA\\
$^{2}$Department of Astronomy  \& Astrophysics, The University of Chicago, Chicago, IL 60637 USA\\
$^{3}$Kavli Institute for Cosmological Physics, The University of Chicago, Chicago, IL 60637 USA\\
$^{4}$Enrico Fermi Institute, The University of Chicago, Chicago, IL 60637 USA
}

\date{Accepted XXX. Received YYY; in original form ZZZ}

\pubyear{2019}

\begin{document}
\label{firstpage}
\pagerange{\pageref{firstpage}--\pageref{lastpage}}
\maketitle

\begin{abstract}
We quantify the shape and overdensity of the galaxy distribution in the `Local Sheet' within a sphere of $R=8$ Mpc, and compare these properties with the expectations of the $\Lambda$CDM model.
We measure ellipsoidal axis ratios of $c/a\approx0.16$ and $b/a\approx0.79$, indicating that the distribution of galaxies in the Local Volume can be approximated by a flattened oblate ellipsoid, consistent with the `sheet'-like configuration noted in previous studies.
In contrast with previous estimates that the Local Sheet has a density close to average, we find that the number density of faint and bright galaxies in the Local Volume is $\approx1.7$ and $\approx5.2$ times denser, respectively, than the mean number density of galaxies of the same luminosity. Comparison with simulations shows that the number density contrasts of bright and faint galaxies within $8$ Mpc alone make the Local Volume a $\approx 2.5\sigma$ outlier in the $\Lambda$CDM cosmology. 
Our results indicate that the cosmic neighbourhood of the Milky Way may be unusual for galaxies of similar luminosity. The impact of the peculiar properties of our neighbourhood on the properties of the Milky Way and other nearby galaxies is not yet understood and warrants further study. 

\end{abstract}

\begin{keywords}
galaxies: abundances -- cosmology: large-scale structure of universe -- catalogues
\end{keywords}



\section{Introduction}
\label{sec:intro}

The distribution and properties of galaxies in the immediate neighbourhood of the Milky Way provide unique and detailed information about the processes driving galaxy formation \citep[e.g.,][]{Peebles.Nusser.2010}.
In particular, the ability to resolve individual stars and to measure detailed properties of the stellar populations and gas distributions of galaxies within a few Mpc  allows for the reconstruction of detailed star formation and chemical enrichment histories  
\citep[e.g.,][]{Weisz.etal.2014}. Additionally, the faintest dwarf galaxies can only be observed to a few tens of kiloparsecs (and to a few hundreds of kpc in the near future), which means that there are entire regimes of galaxy formation that can only be probed locally.

The local $\lesssim8$ Mpc region around the Milky Way -- known as the `Local Volume' (LV) -- is also a unique test bed for cosmological models on small scales. The faintest galaxies are expected to inhabit low-mass dark matter subhaloes whose abundances and internal properties are sensitive to a wide range of cosmological models and to the properties of the dark matter particle \citep[e.g.,][see \citealt{bullock_boylankolchin17} for a review]{Bozek.etal.2016,fitts_etal19}. Additionally, redshift-independent distance measurements in the Local Volume can confirm that satellites are within the virial radii of their hosts, allowing for studies of satellite systems which are free of redshift-space contamination. 

This observational uniqueness of the Milky Way's immediate surroundings justifiably makes it a target of intense modelling efforts. However, to generalise what we learn from the Local Volume, we must either assume that our environment is typical or must be able to model similar environments in simulations. The latter is especially important when there is a good reason to suspect that environment plays a role in particular observational properties.  

The most common approach for defining Milky Way analogues in cosmological simulations is to simply select objects from haloes of similar mass. Some studies require that analogues are in a pair with another halo of a similar mass \citep[e.g.,][]{GarrisonKimmel.etal.2014} and/or that they are sufficiently far from massive groups or clusters \citep[e.g.,][]{Griffen.etal.2016,Geha.etal.2017}. Other studies require that analogues are located within regions with overdensities and local velocity dispersions similar to those estimated for the Milky Way \citep[e.g.,][]{Klypin.etal.2003}. Such criteria are often important, for example, in reproducing dynamical properties of the Local Group \citep{Gonzalez.2014} or for inferring the formation times of nearby galaxies \citep{Forero-Romero.etal.2011}. 

However, the Local Volume has a number of defining properties beyond its density and proximity to clusters. The galaxy distribution within $\lesssim 8$ Mpc around Milky Way is highly flattened in a structure called the `Local Sheet' \citep[][]{Tully.etal.2008} and exhibits coherent motion with a peculiar velocity dispersion of only $\approx 40\,\rm km\,s^{-1}$ \citep{Karachentsev.etal.2003,Klypin.etal.2003,Tully.etal.2008}. The Local Sheet  is embedded in and aligned with a larger flattened structure at scales $\lesssim 40$ Mpc, commonly referred to as the Local Supercluster \citep{DeVaucoulers.1953,DeVaucoulers.1958,Tully.1982}. Theoretically, anisotropy in the large-scale mass distribution is known to correlate with dark matter halo properties \citep[e.g.,][]{Codis.etal.2015,Ramakrishnan.etal.2019}, although the main effect may be due to changes in local overdensity \citep{Goh.etal.2019} and other environmental factors which only correlate with anisotropy \citep{Mansfield.Kravtsov.2019}. The spins of galaxies are expected to correlate with the anisotropy of nearby structures \citep[e.g.,][but see \citealt{Krolewski.etal.2019}]{Navarro.etal.2004,Veena.etal.2019,Kraljic.etal.2019}. Observationally, the orientations of disc galaxies are correlated with the large-scale distribution of their neighbouring galaxies both in the Local Volume \citep{Flin.Godlowski.1986,Flin.Godlowski.1990,Navarro.etal.2004,Noh.Lee.2006} and in the field \citep{Trujillo.etal.2006,Jones.etal.2010,Tempel.etal.2013,Tempel.Libeskind.2013}.  Additionally, \citet{Guo.etal.2015} showed that SDSS galaxies in filaments have $\approx50$ per cent more satellites than similar-luminosity galaxies outside of filaments, further emphasizing the impact that the large-scale environment has on satellite populations.

Another possible impact of the large-scale environment relates to the observed anisotropy of satellite systems.
Although $\Lambda$CDM  generally predicts some level of anisotropy \citep[e.g.,][]{Zentner.etal.2005,Libeskind.etal.2005,Libeskind.etal.2011}, the observed systems exhibit a level of flattening and co-rotation that is  rare in cosmological simulations -- the issue known as the `Planes of Satellites Problem' \citep[see e.g.,][for a recent review]{Pawlowski.2018}. Interestingly, \citet{Libeskind.etal.2015} demonstrated that the satellite distributions around several nearby bright galaxies are aligned with the normal vector of the Local Sheet \citep[see also][]{Libeskind.etal.2019}. The alignment of the observed `satellite planes' with the galaxy distribution in the Local Sheet indicates that our environment likely plays a role in their formation. To study this role using cosmological simulations, one must be able to select environments with properties close to the Local Volume. 

In this regard, it is notable that \citet{Goh.etal.2019} studied the incidence of $\approx 4h^{-1}$ Mpc `walls' in the Bolshoi-Planck $\Lambda$CDM simulation and argued that such walls are exceedingly rare, which would make modelling them challenging. If LV-like environments are truly rare and also significantly influence the properties of the galaxies within them, this could be behind the rarity of the flattened satellite configurations found in $\Lambda$CDM simulations. However, as we discuss in Section \ref{sec:comparison}, the apparent rarity of LV analogues is in large part due to the specific and restrictive choices used by these authors to define LV-like environments. Nevertheless, as we will show in this study, even basic properties of the galaxy distribution in the Local Volume, such as the abundance of bright galaxies relative to faint galaxies, make it a $\approx 2.5\sigma$ outlier compared to the expectations of $\Lambda$CDM cosmology.  

The Local Volume contains a large number of bright and massive galaxies \citep{McCall.2014,Karachentsev.Kudrya.2014,Klypin.etal.2015,Kourkchi.Tully.2017}. The ring-like concentration of bright galaxies at distances of $\approx 3-5$ Mpc around the Local Group was dubbed `The Council of Giants' by \citet{McCall.2014}. This means that some isolation criteria that are commonly used when defining Milky Way analogues in simulations will preferentially select environments systematically less crowded than that of the Milky Way. The impact of this difference is not yet fully understood. 

These considerations illustrate that it is important to understand the connection between the Milky Way's properties and its environment, and that it may be crucial to match these properties in simulations aiming to reproduce the properties of the Milky Way and its neighbours. {\em However, to do this we must understand what the properties of the Local Volume are}. Thus, in this study we characterise basic properties of the Local Volume and estimate how common such environments are in $\Lambda$CDM cosmological simulations. 

Although there are many different ways to define the local environment and its properties, we adopt a simple approach that allows us to capture and estimate two key features of the Local Volume: the overdensity of galaxies at various luminosities, and the flattened shape of the galaxy distribution. 

To this end, we use the \citet{Karachentsev.etal.2013} local galaxy catalogue after making some important updates to the $B$-band apparent magnitudes, dust extinction corrections, and distances (Section \ref{sec:local}). We use the Millennium Galaxy Catalogue \citep[see][and Section \ref{sec:mgc} below]{Liske.etal.2003,Driver.etal.2005,Allen.etal.2006} to estimate the mean density of galaxies in the field. We use these galaxy catalogues to characterise the shape and density contrast of the galaxy distribution in the Local Volume in Section \ref{sec:characterising}. In Section \ref{sec:unusual} we compare our results against the predictions of the SMDPL simulation of \citet{Klypin.etal.2016} using the methodology described in Section \ref{sec:methods}. We discuss our results in Section~\ref{sec:discussion}. We summarise our results and conclusions in Section~\ref{sec:conclusions}.

\section{Galaxy samples in the local volume and in the field}

\begin{table*}
  \centering
  \caption{Ten randomly selected galaxies from our augmentation of the LVG, ordered by distance. The columns shown here are the galaxy name, the Vega asymptotic $B$-band magnitude, $B_T$, the $B$-band galactic extinction, $A_{B,{\rm gal}}$, the $B$-band internal extinction, $A_{B,{\rm int}}$, the distance modulus to that galaxy, the method used to determine photometry, and the method used to determine the galaxy's distance. The $B_T$ method code corresponds to the numbered items in Section \ref{sec:photometry}. The error bars on the distance moduli represent only the statistical errors on our best-fitting distance moduli, $\mu_\star$; we discuss additional sources of errors at length in Section \ref{sec:distances}. The electronic version of this table also includes ADS bibliographic codes for source papers and reproduces supergalctic coordinates and Holmberg radii from \citet{Karachentsev.etal.2013}. The electronic version of this table will be made public upon request.
  }
  \label{tab:augment}
  \begin{tabular}{lllllll}
  \hline\hline
    Galaxy Name & $B_T$ & $A_{B,{\rm gal}}$ & $A_{B,{\rm int}}$ & DM & $B_T$ Method & Preferred DM Method \\
  \hline
    LMC & 0.91 & 0.32 & 0.02 & $18.48\pm0.01$ & 1 & Eclipsing Binary \\
    MESSIER032 & 9.03 & 0.26 & 0.18 & $24.21\pm0.12$ & 1 & TRGB \\
    NGC2403 & 8.93 & 0.18 & 0.33 & $27.51\pm0.02$ & 1 & TRGB \\
    NGC4945 & 9.30 & 0.76 & 1.16 & $27.73\pm0.04$ & 2 & TRGB  \\
    ESO223-009 & 13.82 & 1.12 & 0.02 & $28.19\pm0.03$  & 3 &  TRGB \\
    ESO273-014 & 12.90 & 1.32 & 0.17 & $29.05\pm0.25$  & 1 & Tully--Fisher \\
    HIJASS J1021+6842 & --- & 0.09 & ---  & $29.19$ & --- & Membership \\
    NGC4656 & 10.96 & 0.06 & 0.44 & $29.48\pm0.37$ & 1 & TRGB \\
    DDO217 & 12.78 & 0.62 & 0.11 & $29.83\pm0.40$ & 1 & Tully--Fisher \\
    NGC2784 & 11.30 & 0.93 & 0.00 & $29.92\pm0.12$  & 1 &  SBF \\
  \hline
  \end{tabular}
\end{table*}

\begin{table}
  \centering
  \caption{The median intrinsic scatter, $\sigma_\star$, for different {\em unhomogenised} distance measurement methods, taken from galaxies with a large number of measurements of each type. These values are described in detail in Section \ref{sec:distances}. These values only estimate the typical impact of not homogenising distance measurements and cannot be interpreted as the systematic errors of these measurement methods.}
  \label{tab:intrinsic}
  \begin{tabular}{ll}
  \hline\hline
    Method & $\sigma_\star$ (mag) \\
  \hline
    Eclipsing Binary & 0 \\
    TRGB & 0.062 \\
    Cepheid & 0.099 \\
    SBF & 0.092 \\
    Tully--Fisher & 0.29 \\
  \hline
  \end{tabular}
\end{table}

We use the most extensive catalogue of local galaxies that has been compiled to date -- The Catalog and Atlas of the Local Volume Galaxies \citep[the LVG,][]{Karachentsev.etal.2013}.\footnote{\href{http://www.sao.ru/lv/lvgdb/}{\texttt{http://www.sao.ru/lv/lvgdb}}} We augmented the available Johnson $B$-band magnitude estimates, the internal extinction corrections, and the distances for a subset of the LVG galaxies. We describe this process in detail in Section \ref{sec:local}.

To estimate the overdensity of galaxies of different luminosities in the Local Volume, we need a reference field luminosity function. There are many spectroscopic surveys which have measured luminosity functions at low redshifts, but for a comparison with the LVG, a survey should collect photometry in a band which can be easily converted to the Johnson $B$-band magnitudes used by the LVG, it should allow for the measurement of the 26.5 mag arcsec$^{-2}$ Holmberg radius, and it should allow for internal extinction corrections. Finally, these survey data should be publicly available so that conversions, corrections, and cuts can be performed directly on individual galaxies. 

The Millennium Galaxy Catalogue  \citep[MGC,][]{Liske.etal.2003,Driver.etal.2005,Allen.etal.2006} best matches these criteria. We describe the MGC in Section \ref{sec:mgc} along with the choices we make in sample selection, internal extinction corrections, and surface brightness calculations. We also describe how we estimate the luminosity function of the MGC galaxies, using a surface brightness limit, in Section \ref{sec:mgclf}.

\subsection{The Local Volume}
\label{sec:local}

To study the distribution of galaxies in the Local Volume we use the LVG  \citep{Karachentsev.etal.2013}. The LVG is a galaxy catalogue that compiles asymptotic $B$- and $K_s$-band magnitudes, extinction corrections, distances, and a range of other properties for all known galaxies within 11 Mpc from the Milky Way. It is continuously updated as new galaxies are discovered. Below, we describe how we use this catalogue as well as our updates to its $B$-band magnitudes, distances, and extinction corrections using available information from the literature. 

\subsubsection{Photometry}
\label{sec:photometry}

The first choice we must make is between using optical $B$-band magnitudes and near infrared (NIR) $K_s$ magnitudes. NIR measurements would normally be preferable because they are less affected by galactic and internal extinction, and they are expected to correlate with stellar mass more tightly than optical luminosities \citep[e.g.,][and references therein]{McGaugh_Schombert.2014}. Furthermore, the existence of the 2MASS all-sky NIR survey \citep{Skrutskie.etal.2006} would mean that all galaxies would have photometry which was collected with uniform methodology. Unfortunately, as noted by \citet{Kirby.etal.2008} and \citet{McCall.2014}, the 2MASS survey can underestimate absolute magnitudes of nearby galaxies by up to 2.5 mag in the $K_s$ band, often encountering problems with low surface brightness galaxies, galaxies that are large and bright, and galaxies near the edges of survey strips. We find similarly significant discrepancies when comparing nearby 2MASS magnitudes against the high-quality NIR measurements in \citet{McGaugh_Schombert.2014}, even when applying identical internal and galactic extinction corrections. Given that the 2MASS survey is the dominant source of NIR photometry in the Local Volume, these issues motivate us to use $B$-band magnitudes.

$B$-band magnitudes pose their own problems. The main issue is the inhomogeneity of these measurements, which have been collected over almost a century by various authors using data of highly variable quality and different methodologies. The LVG catalogue reports a $B$-band magnitude for almost every galaxy, but in many cases only very old magnitude estimates are available (e.g., \citealp{Ames.1930} magnitudes reported via \citealp{RC3}). In cases where no extant measurements could be found, the LVG authors report visual estimates of $B$-band magnitudes by comparing with galaxies that have qualitatively similar structures.

The most common source for $B$-band magnitudes in the Local Volume is the Third Reference Catalog for Bright Galaxies \citep[RC3;][]{RC3}. The RC3 homogenises photometry of most galaxies into relatively high-quality Vega-magnitude $B_T$ magnitudes by extrapolating surface brightness profiles to total asymptotic magnitudes with morphology-dependent growth curves. However, for a subset of galaxies in the RC3, the only existing magnitudes came from much older, pre-CCD photographic surveys
\citep[e.g.][]{Ames.1930,Shapley_Ames.1932,CGCG}. Although RC3 provides average conversions for these magnitudes to the $B_T$ system, all such measurements -- labelled $m_B$ -- have large ($\sim 0.3-0.5$ mag) systematic and statistical errors.

To improve this, we update the LVG catalogue by performing an additional literature review on the brightest nearby galaxies with fiducial LVG distances smaller than 10 Mpc and fiducial LVG corrected $B$-band magnitudes brighter than $-15$ or which did not have magnitude entries in the LVG. We also consider galaxies which do not have magnitude entries in the LVG. This gives us a starting sample of 312 galaxies. First, we identify the highest quality $B$-band magnitudes available for each galaxy using the following multi-step procedure:
\begin{enumerate}
    \item We take the $B_T$ entries from RC3, where available. This was done for 171 galaxies.
    \item For the galaxies remaining after step (i), we search the photometric surveys used by \citet{Karachentsev.etal.2013} along with an additional ten $B$-band surveys. In the cases where multiple surveys report magnitudes, we take the value with the lowest estimated error. This was done for 68 galaxies.
    \item For the galaxies remaining after step (ii), we use NED\footnote{The NASA/IPAC Extragalactic Database (NED) is funded by the National Aeronautics and Space Administration and operated by the California Institute of Technology.}, HyperLeda\footnote{\href{http://leda.univ-lyon1.fr}{\texttt{http://leda.univ-lyon1.fr}}} \citep{Makarov.etal.2014}, and Simbad \citep{Wenger.etal.2000} to search for asymptotic $B$-band magnitudes. We also perform extensive literature reviews on each galaxy in this group to identify magnitude measurements not placed in one of these databases. This was done for 26 galaxies.
    \item For the galaxies remaining after step (iii), we assign magnitudes from the $m_B$ values in the RC3 or other pre-CCD catalogues. This was done for 16 galaxies.
    \item For the galaxies remaining after step (iv), we identify galaxies with asymptotic $ugriz$ magnitudes using a procedure similar to steps (ii) and (iii) and convert them to $B$-band magnitudes through the relations in \citet{Blanton.Roweis.2007}. If available, we prioritise using the $g$ and $r$ bands to do this. This was done for 5 galaxies.
    \item For the galaxies remaining after step (v), we use visually-estimated LVG magnitudes where available. This was done for 16 galaxies.
    \item For the galaxies remaining after step (vi), we use magnitudes from \citet{Paturel.etal.2000}. This was done for 2 galaxies.
    \item There were eight additional galaxies which we were unable to find visual magnitudes for.
\end{enumerate}

At the end of this process, we are left with seven galaxies which were identified from HI surveys but for which no optical magnitudes are available in the LVG or in the literature.  Several of these galaxies are within the SDSS footprint. We visually inspected these fields and found that some of these galaxies were coincident with bright foreground sources, but for most we could not identify an optical counterpart. We were able to easily identify optical counterparts for galaxies slightly below our surface brightness cutoff (see Section \ref{sec:mgc}) using the same procedure, so this indicates that these galaxies would likely be removed by this cut anyway.

The updated total magnitudes, $B_T$, are provided in a supplementary data table, along with citations. Example entries are shown in Table \ref{tab:augment}. In all cases, we manually reverse any internal or external extinction corrections applied in the original analysis, so that we can apply these corrections uniformly to the entire sample and convert AB magnitudes to Vega magnitudes (where applicable) using the relations in \citet{Willmer.2018}.

\subsubsection{Extinction}
\label{sec:extinction}

\begin{figure}
    \centering
    \includegraphics[width=0.47\textwidth]{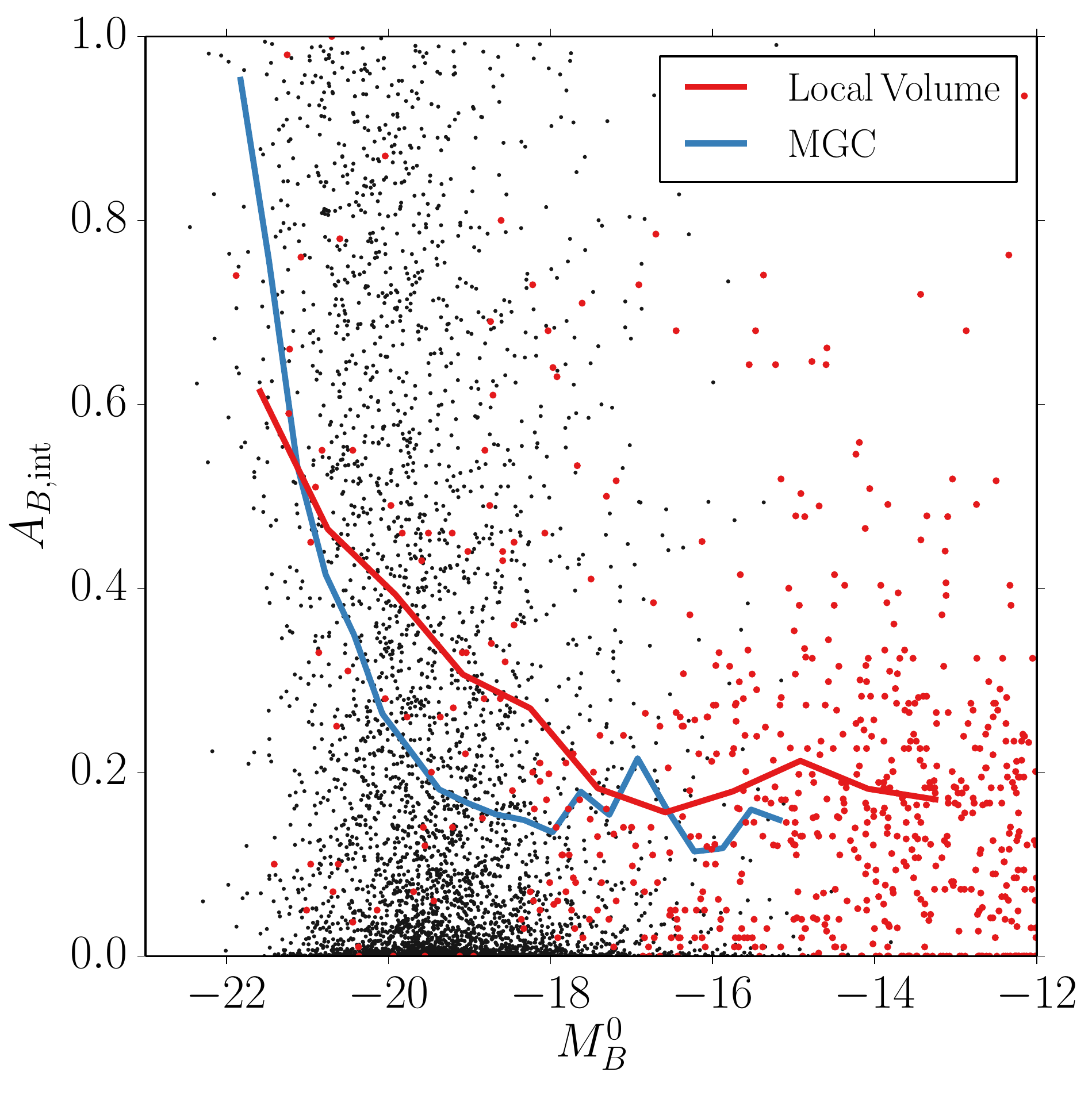}
    \caption{Comparison between MGC internal extinction corrections \citep[Section \ref{sec:mgc}, see also][]{Driver.etal.2007} and the local internal extinction corrections used in this paper (see Section \ref{sec:local}). MGC galaxies are shown as black points and local galaxies are shown as red points. The medians of these distributions are shown as blue and red curves, respectively. Only late-type galaxies are shown because internal extinction depends strongly on morphology and the Local Volume has a different morphology distribution from the large-scale average.  The two approaches agree reasonably well at the dim end and for bright galaxies with $M_B^0 \lesssim -20.5$, the two most important regimes for our analysis.}
    \label{fig:internal_extinction}
\end{figure}

\begin{figure}
    \centering
    \includegraphics[width=0.475\textwidth]{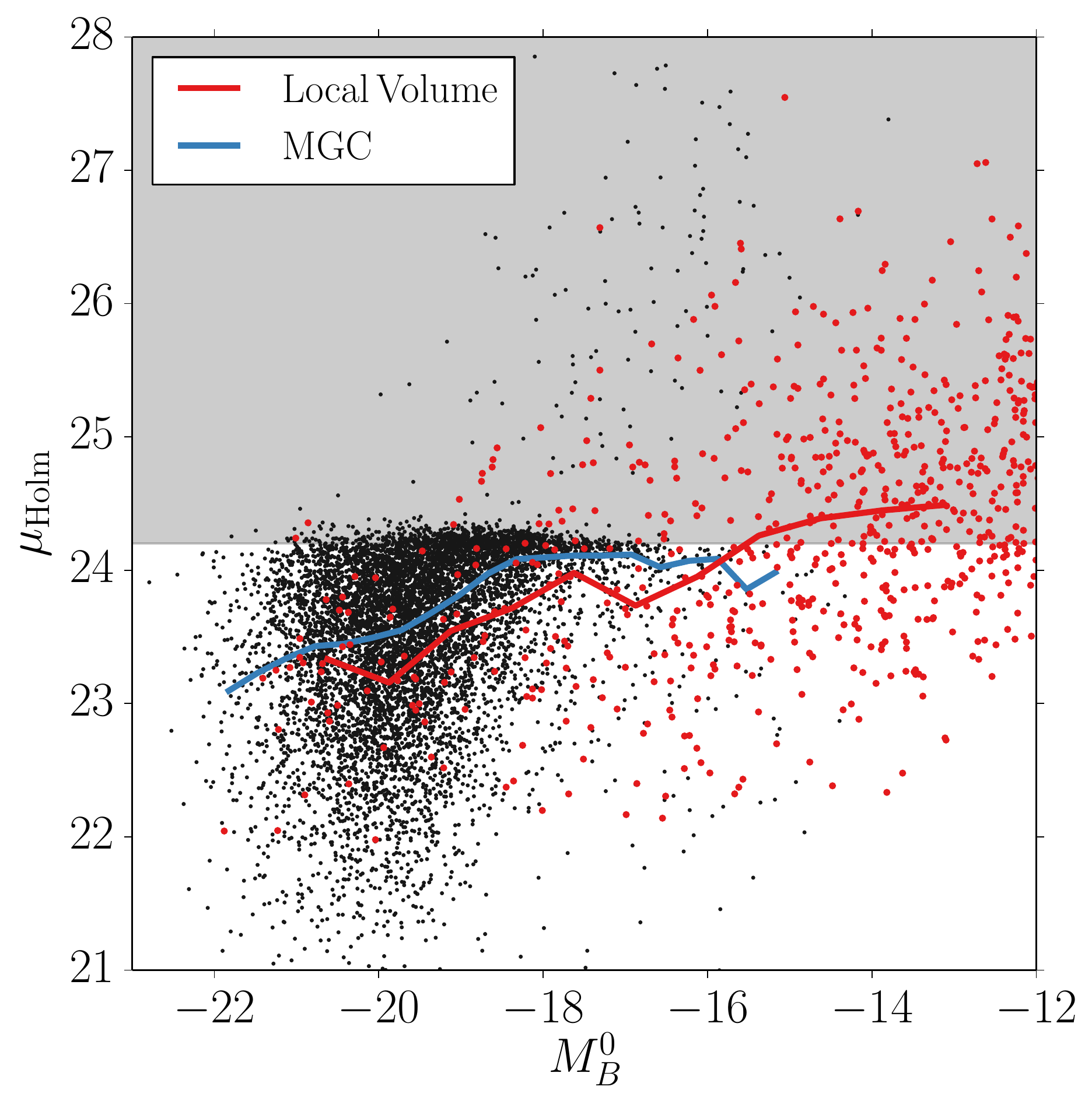}
    \caption{The distribution of surface brightnesses in the MGC spectroscopic survey and in the Local Volume relative to the 26.5 mag arcsec$^{-2}$ Holmberg radius. This plot uses the same colour scheme as Fig.~\ref{fig:internal_extinction}. The MGC becomes incomplete at $\mu_{\rm Holm} \gtrsim 24.2.$ A significant fraction of local galaxies are measured at dimmer surface brightnesses than this limit, necessitating that a comparable cut be made to this sample. At bright magnitudes the two samples have qualitatively similar $\mu_{\rm Holm}-M_B^0$ trends. Note that some of the lowest surface brightness galaxies in the LVG use an alternative definition of surface brightness (see Section \ref{sec:surface_brightness}). We do not use any of these galaxies.}
    \label{fig:surface_brightness}
\end{figure}

Galaxy light is extinguished when it passes through dust in both the source galaxy and in the Milky Way. We perform all our analysis on magnitudes that are corrected for this extinction:
\begin{equation}
    B_T^0 = B_T - A_{B,{\rm gal}} - A_{B,{\rm int}}.
\end{equation}
Here, $A_{B,{\rm gal}}$ is the $B$-band galactic extinction which we estimate from the dust maps of \citet{Schlegel.etal.1998} and $A_{B,{\rm int}}$ is the internal extinction due to dust in the source galaxy.

We use the internal extinction correction method described in \citet{Karachentsev.etal.2013}, which is based on the HI line-width method of \citet{Verheijen.etal.2001}. For galaxies without HI measurements, we use the morphology-based internal extinction scheme from \cite{RC3}. The \cite{RC3} correction is parametrized as
\begin{equation}
A_{B,{\rm int}} =
\begin{cases} 
       1.5 - 0.03(T - 5)^2 \log_{10}(b/a) & :\,\,\,T_{\rm Hubble} \geq 0 \\
      0 & :\,\,\, T_{\rm Hubble} < 0
   \end{cases}
\end{equation}
where $T_{\rm Hubble}$ is the galaxy's numerical Hubble stage according to the classification system adopted in RC3 and $b/a$ is the on-sky axis ratio of that galaxy. {\color{black} The internal extinction corrections applied to the Local Volume are shown in Fig.~\ref{fig:internal_extinction}.}

\subsubsection{Distances}
\label{sec:distances}

Rather than manually picking a single distance estimate for each galaxy, we choose the most reliable distance measurement method available for each galaxy and combine all measurements with that method. In the cases of highly discrepant independent distance estimates, we select the most reliable one after reviewing details of the measurements.

First, for each galaxy we collect all distance measurements with estimated uncertainties available in NED and group these measurements by method. Then, we fit a Gaussian distribution to these measurements by maximizing the likelihood function
\begin{equation}
    \ln{(\mathcal{L}(\mu_\star, \sigma^2_\star))} = -\sum_i^N\frac{1}{2}\ln{(2\pi(\sigma_\star^2 + \sigma_i^2))} + \frac{(\mu_\star - \mu_i)^2}{2 (\sigma_\star^2 + \sigma_i^2)}.
\end{equation}
Here, $\mu_\star$ and $\sigma^2_\star$ are the mean and variance of the fitted distribution, respectively. $\mu_i$ and $\sigma_i$ are the distance and error for each measurement. 

The fit is performed in distance modulus space. At the conclusion of this step, every galaxy has distance estimates, $\mu_\star$, for a variety of distance methods as well as estimates of the `intrinsic' scatter, $\sigma^2_\star$, needed to explain the diversity in measurements of that type. For galaxies with measurement errors consistent with the scatter between measurements, $\sigma^2_\star$ will be zero.

We do not manually homogenise measurements (as done, for example, by \citealp{Freedman.etal.2019}) due to the large number of galaxies that we analyse. This will manifest itself as a larger dispersion in measurements than would be expected from the reported errors, and thus non-zero values of $\sigma_\star.$ To estimate the impact of this, Table \ref{tab:intrinsic} presents the median $\sigma_\star$ values for for each method using only galaxies with  at least 5 independent distance measurements and shows that $\sigma_\star$ is $\lesssim0.1$ mag, except for Tully--Fisher-based distance measurements, which reach $\sigma_\star=0.29$ mag. Analysis which requires more precise distances than these must manually homogenise measurements. Note that these values {\em cannot} be interpreted as systematic errors of these methods: they combine both modern high-quality measurements with older measurements with poorly understood systematic uncertainties. Additionally, if many measurements with the same underlying assumptions are made of the same galaxy, that galaxy will have a small $\sigma_\star$, regardless of the true systematic uncertainty. Rather, the values of $\sigma_\star$ in Table \ref{tab:intrinsic} represent only the global measurement uncertainty due to inhomogeneity in the method assumptions used to make them. 

Next, we adopt $\mu_\star$ from the method with the lowest uncertainty on the distance modulus. For simplicity we take this uncertainty to be the statistical error on $\mu_\star$ added in quadrature with the median $\sigma_\star$ from Table \ref{tab:intrinsic}. This almost always prioritises eclipsing binary distances over Cepheid and TRGB distances, prioritises the latter over surface brightness fluctuation distances, and assigns Tully--Fisher distances the lowest priority. Studies that rely on the numeric values of these errors will need to perform more detailed analysis. 

A small number of galaxies have no distance measurements in NED. In these cases, we manually select the highest-quality distance measurement that we can find using both the LVG distance estimates and our own literature review. This sometimes requires that distances are reported based on assuming membership in a nearby group from redshift measurements. About 4 per cent of galaxies within 10 Mpc have membership-based distances. 

Once we have assembled the distances for our entire sample, we use these distances to compute absolute magnitudes,
\begin{equation}
    M_B^0 = B_T^0 - DM.
\end{equation}
As a test, we cross-match our sample against the the high quality distance measurements in \citet{Freedman.etal.2019}. We find that our automated method reproduces their distance to the LMC to within 0.01 mag and that the rms offset between our TRGB distances and the \citet{Freedman.etal.2019} TRGB distances is 0.11 mag. This number is consistent with the typical uncertainty that we report on our unhomogenised TRGB distances for this sample. While our method should not be used for applications which need either extreme precision or thoroughly characterised errors, it produces distances which are acceptable for the purposes of this study.

After performing this analysis, we identify galaxies which have particularly large intrinsic scatter, $\sigma_\star$, relative to reported errors and galaxies which are bright but only have low-quality distance measurements on NED. For these galaxies, we inspected individual distance estimates and performed additional literature reviews to find distances not reported on NED. Of particular note is the Maffei group -- a (possible) group of three bright galaxies which are near the Milky Way, but are heavily obscured by the Milky Way's disc. We use distances and extinction measurements from \citet{Anand.etal.2019} for Maffei 1, Maffei 2 and IC0342. We choose these measurements due to the careful analysis of galactic extinction and the corrections that are performed on previous distance measurements.

We include these combined distances along with errors and citations in a supplementary data table. Several random entries from this table are {\color{black} also} given in Table \ref{tab:augment}.

\subsubsection{Surface Brightness Estimates}
\label{sec:surface_brightness}

Surface brightness profiles are not available for many LVG galaxies. The catalogue provides Holmberg radii corresponding to the the $26.5$ mag arcsec$^{-2}$ isophote, which can be combined with $M_B^0$ to obtain the corresponding surface brightness, $\mu_{\rm Holm}$. Some dwarf galaxies have such low surface brightnesses that measurement of the $26.5$ mag arcsec$^{-2}$ isophote is not feasible or even possible. In these cases, the LVG reports the galaxy's effective radius as its Holmberg radius. This means galaxies with low surface brightnesses may use either radius definition without indication, complicating the evaluation of the associated uncertainty. However, as we describe in Section \ref{sec:mgc}, we do not use galaxies near this surface brightness limit in our main analysis, although they do appear in Fig.~\ref{fig:surface_brightness}.

\subsection{The Millennium Galaxy Catalogue as a Field Galaxy Sample}
\label{sec:mgc}

\begin{figure}
    \centering
    \includegraphics[width=0.475\textwidth]{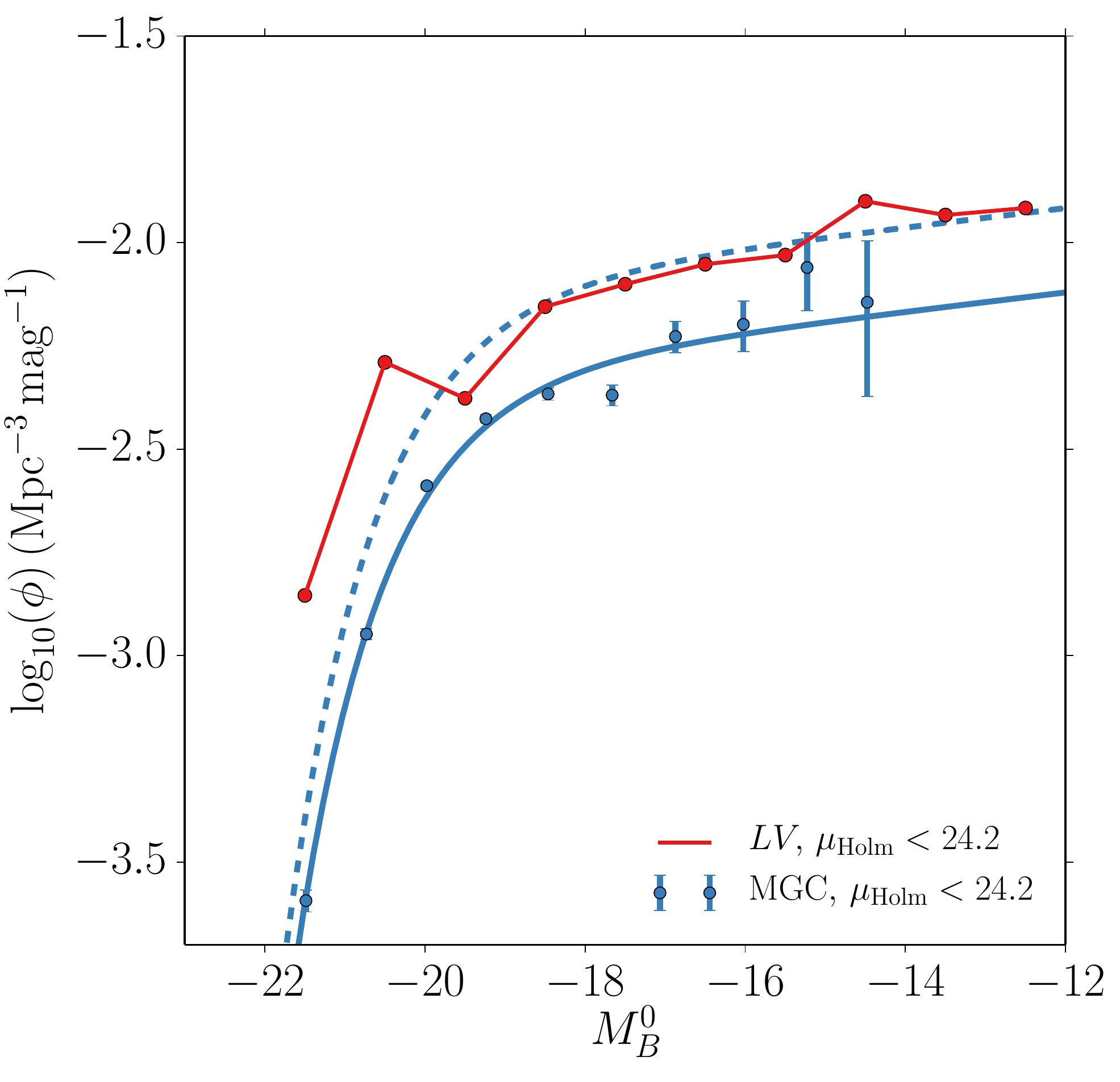}
    \caption{Comparison of the luminosity functions measured in the Local Volume and in the MGC spectroscopic survey. The red curve shows the luminosity function within 8 Mpc after a cut on $\mu_{\rm Holm}$ has been made to match the MGC cutoff. The blue points show the MGC luminosity function, the blue curve shows the best-fitting Schechter function, and the dashed blue line shows this luminosity function with a multiplicative offset of 1.66 to match the faint end amplitude of local luminosity function. Note that this fit is provided purely for illustrative purposes: we compare against the MGC non-parametrically throughout this paper.}
    \label{fig:lum_func}
\end{figure}

The MGC is a 37.5 deg$^2$ $B$-band galaxy survey \citep{Liske.etal.2003} which is 96 per cent complete down to $B < 20$ and 99.6 per cent complete down to $B < 19$ \citep{Driver.etal.2005}, allowing it to probe the luminosity function down to $M_B \lesssim -14.5$.

The MGC $B$-band is slightly different from the Johnson $B$-band used in the LVG catalogue, so we perform a conversion on all MGC magnitudes prior to analysis. For a galaxy with known  $B-V$ colour we convert between the two using the relation
\begin{equation}
    B = B_{\rm MGC}  - 0.145(B-V)
\end{equation}
\citep{Cross.etal.2014}. Reliable colours are not available for all galaxies in the MGC or the LVG catalogue, so we apply a bulk correction to the entire MGC sample. We find that the mean $B-V$ colour of the LVG galaxies within 8 Mpc that have extant $V$-band measurements is 0.69 mag with a standard deviation of 0.22 mag, corresponding to a mean colour-dependent shift of 0.10 $\pm$ 0.03 mag. This colour-dependent uncertainty is subdominant to the per-galaxy error in the MGC's zero-point correction of 0.09 mag \citep{Cross.etal.2014}, and thus a more detailed correction is not necessary.

\citet{Driver.etal.2007} derived an empirical internal extinction correction for the MGC by requiring that galaxies with different inclinations fall along identical luminosity functions. They performed separate fits for the extinction in the disc and bulge light components:
\begin{align}
    A_{B,{\rm int},{\rm disc}} &= 0.99\, (1 - \cos{i})^{2.32} \\
    A_{B,{\rm int},{\rm bulge}} &= 2.16\, (1 - \cos{i})^{2.48},
\end{align}
where $i$ is the inclination of the disc. For a known bulge-to-total light ratio, $B/T$, this can be converted into a total internal extinction correction, $A_{B,{\rm int}}$.

The bulge-to-total light decomposition is not uniformly available across the LVG, just as HI line widths and detailed morphological information are not available for MGC galaxies. This means that the same internal extinction correction scheme cannot be used for both samples. We compare $A_B$ for our two samples using these separate correction methods in Fig.~\ref{fig:internal_extinction}. We find good agreement at both very low and very high magnitudes, and a slight disagreement of $\approx 0.15$ mag for intermediate magnitudes in the range $-20 < M_B^0 < -18.$ Thus, we do not expect any issues arising from using different internal extinction corrections in the LVG and MGC samples. 

Surface brightnesses in the LVG are computed using the 26.5 mag arcsec$^{-2}$ Holmberg radius, a quantity which is not directly available in the MGC catalogue. We estimate the Holmberg radius for each galaxy in the MGC by using the GIM2D \citep{Simard.etal.2002} fitting parameters provided by \citet{Allen.etal.2006}. We compare the distribution of surface brightnesses between the MGC and LVG in Fig.~\ref{fig:surface_brightness}, which shows that the MGC has a surface brightness completeness limit of $\mu_{\rm Holm}\approx 24.2.$

\subsubsection{Constructing the MGC Luminosity Function}
\label{sec:mgclf}

To estimate the luminosity function (LF) of the MGC, we first remove galaxies with surface brightnesses above the MGC limit ($\mu_{\rm Holm} > 24.2$), galaxies outside the redshift range $0.013 < z < 0.18$, all galaxies with $B_{\rm MGC,AB} > 20$, and galaxies with GIM2D fitting errors which led to bulge fractions $\geq 1$ or excessively large S\'ersic indices, $n$. We found that a cut of $n < 90$ removes all such cases.

To correct for Malmqiust bias, we compute the effective volume within which a galaxy could have been observed in the MGC:
\begin{align}
    V_{\rm eff} &= f \frac{4\pi}{3}\left[{\rm min}\left(R_{\rm max},\,R_{\rm lim}(B)\right)^3- R_{\rm min}^3\right] \\
    f &= f_\Omega\cdot f_{\rm spec} \cdot f_{\rm GIM2D}.
\end{align}
Here, $R_{\rm min}$ and $R_{\rm max}$ are the comoving distances corresponding to $z=0.013$ and 0.18, respectively, $R_{\rm lim}(B)$ is the maximum comoving radius at which a galaxy with apparent magnitude $B$ could be detected within the $B_{\rm MGC,AB} = 20$ limit. $f_\Omega = 7.48\times 10^{-4}$ corrects for the survey area of the MGC, $f_{\rm spec} = 0.96$ corrects for the spectroscopic incompleteness at our chosen magnitude limit, and $f_{\rm GIM2D} =0.985$ corrects for galaxies with improper GIM2D fits. Strictly speaking, spectroscopic completeness and GIM2D reliability are functions of apparent magnitude, but we do not model this due to the small size of the effect.

Finally, we construct a non-parametric cumulative luminosity function by first computing the absolute magnitude
\begin{equation}
    M^0_B = B - A_{B,{\rm gal}} - A_{B,{\rm int}} - DM - K + 2.5\times 0.75 \log_{10}(1+z).
\end{equation}
Here, $A_{B,{\rm gal}}$ is the galactic extinction correction described in \citet{Liske.etal.2003}, $A_{B,{\rm int}}$ is the internal extinction correction described above, $K$ is the $K$-correction described in \citet{Driver.etal.2007}, DM is the distance modulus to the galaxy assuming the same cosmology as our simulations (see Section \ref{sec:sim}), and the last term is the evolution correction from \citet{Phillipps.Driver.1995}. Next, we place absolute magnitudes into a very finely binned histogram weighted by 1/$V_{\rm eff}$ and compute the cumulative sum to get the cumulative luminosity function.

To determine the completeness limit of the estimated cumulative luminosity function, we compare against the much deeper LVG. We inspect the ratio of $\langle N_{8,{\rm MGC}}(<M_B^0)\rangle/N_{8,{\rm LVG}}(<M_B^0)$ as a function of $M_B^0$ after removing all galaxies with $\mu_{\rm Holm} > 24.2$ from both samples. Here, $\langle N_{8,{\rm MGC}}(<M_B^0)\rangle$ is the average number of galaxies brighter than $M_B^0$ in spheres of $8$ Mpc radius estimated from the MGC luminosity function, and $N_{8,{\rm LVG}}(<M_B^0)$ is the number of LVG galaxies brighter than $M_B^0$ within 8 Mpc of the Milky Way. We find that the ratio is roughly constant from $-18 < M_B^0 < -15$, and then starts to rapidly decrease. We conclude that the cumulative luminosity function is robust for $M_B^0\lesssim -15$. 

Although we will compare against the non-parametric estimate of the cumulative MGC luminosity function throughout this paper, for reference we provide the best-fitting parameters for its differential form using the \citet{Schechter.1976} approximation
\begin{align}
    \phi(M_B^0) = \frac{\ln{10}}{2.5}\,\phi^\star \left[10^{0.4(M^\star - M_B^0)}\right]^{\alpha+1}\exp\left[-10^{0.4(M^\star - M_B^0)}\right].
\end{align}
For the sample with the surface brightness cut of $\mu_{\rm Holm} < 24.2$, we find $\phi^\star=5.270$ Mpc$^{-3}$ mag$^{-1}$, $M^\star=-20.34$, and $\alpha=-1.058.$ We compare this fit and the underlying MGC data to the luminosity function of the Local Volume in Fig.~\ref{fig:lum_func}.

We find that the faint end slope of the Local Volume is identical to the MGC, although both samples have slopes which change drastically with surface brightness cutoff for $M_B^0 \gtrsim - 19$ \citep[as also shown by][]{Driver.etal.2007}, so this should not be interpreted as the `true' faint end slope, and rather as the slope of the subset of galaxies defined by our surface brightness cut.

\section{$\Lambda$CDM Simulation and Modelling Methods}
\label{sec:methods}

\subsection{The SMDPL Simulation}
\label{sec:sim}

We use the SMDPL cosmological simulation \citep{Klypin.etal.2016} which followed the evolution of $3840^3$ particles in a periodic cube with comoving volume ($400\ h^{-1}$ Mpc)$^3$ assuming cosmological parameters of $h=H_0/100 = 0.678$, $\Omega_M=0.307$, and $\sigma_8=0.829$, consistent with constraints from the Planck observatory. SMDPL was run using the \textsc{L-Gadget-2} $N$-body code, a version of the \textsc{GADGET-2} code optimised for memory efficiency \citep{Springel.2005}. This simulations uses a Plummer-equivalent force softening scale of 1.5 $h^{-1}$kpc and a timestepping parameter of $\eta=0.01$; for the number of particles and the adopted $\Omega_M$, the particle mass is $m_p=9.6\times10^7$ $h^{-1}M_\odot.$

We perform our analysis using \textsc{Rockstar} halo catalogues with \textsc{consistent-trees} post-processing \citep{Behroozi.etal.2013a,Behroozi.etal.2013b}, which was first presented for this simulation in \citet{RodriguezPeubla.etal.2016}. 

Throughout this paper we adopt the primary mass definition used by these catalogues, the overdensity mass $M_{\rm vir}$. $M_{\rm vir}$ is the bound mass enclosed by the radius corresponding to a density contrast of $\Delta_{\rm vir}$ \citep{Bryan.Norman.1998}. We also use the maximum mass attained by each halo across its mass accretion history, $M_{\rm peak}$, to assign absolute magnitudes to haloes using abundance matching. 

Our analysis often requires finding haloes located within a given sphere. Following a commonly used approach, we optimise this operation by first allocating the haloes to cells in a 3D grid spanning the simulation box. To determine which haloes lie within a given sphere, we first determine which grid cells intersect this sphere and retrieve only the haloes contained in those cells for distance calculations (see appendix B of \citealp{Mansfield.Kravtsov.2019} for further discussion).

{\color{black}To aid analysis of the large SMDPL halo catalogues, we have compressed all halo catalogues using a new  compression algorithm, implemented in the \textsc{Minnow}  code \citep{Mansfield.in.prep}. This reduces catalogue sizes by a factor of 10 without the removal of any haloes and decreases read times by a factor of $10^3.$}

\subsection{Luminosity Assignment using Abundance Matching}
\label{sec:abundance_matching}

To assign luminosities to simulated dark matter haloes, we abundance match the MGC luminosity function, estimated as described above, against the SMDPL $M_{\rm peak}$ mass function. Specifically, we derive a mean $M_{\rm peak}-M_B^0$ relation by abundance matching both distinct haloes and subhaloes, and we account for the scatter in the relation using the approach outlined in \citet{Kravtsov.2018}. We experimented with the impact of different assumptions about the scatter in the $M_{\rm peak}-M_B^0$ relation and found that scatter has a negligible effect on our results. Therefore, to assign luminosities we simply evaluate the mean relation at a given halo's $M_{\rm peak}$. {\color{black} Based on convergence testing that we are preparing for publication, we required that haloes have 800 particles for valid $M_{\rm peak}$ measurements, meaning that our abundance matching scheme is valid within the range $\log_{10}(M_{\rm peak})=10.91$ ($B=-16.03$) to $\log_{10}(M_{\rm peak})=14.38$ ($B=-22.30$).}

Although abundance matching using $M_{\rm peak}$ instead of a current-epoch mass allows us to sensibly assign luminosities to subhaloes, \citet{Campbell.etal.2018} show that abundance matching using $V_{\rm peak}$ instead of $M_{\rm peak}$ leads to satellite fractions and 5 Mpc correlation functions which are approximately $10$ per cent more accurate at low stellar masses. We chose to use $M_{\rm peak}$ because the convergence testing results of  \citet{Mansfield.Avestruz.in_prep} show that $M_{\rm peak}$ is converged to lower resolutions than $V_{\rm peak}$, allowing us to reach abundances similar to the Local Volume at its completeness limit.

Although the MGC has a deeper surface brightness cutoff than many other redshift surveys (see comparison with SDSS and 2dFGRS in \citealp{Cross.etal.2014}), Fig.~\ref{fig:surface_brightness} shows that the surface brightness limits of the MGC would begin to miss local galaxies at $M_B^0 \gtrsim -19$, in agreement with fig.~16 in \citet{Driver.etal.2005}.\footnote{As discussed in Section \ref{sec:mgc}, the MGC uses a different surface brightness definition, so this comparison must be done by matching faint-end slopes.} This means that for $M_B^0 \gtrsim -19$, abundance matching will populate dark matter haloes with underestimated luminosities.

In principle, one solution is to assume a distribution and extrapolate outside the complete regions for the MGC and Local Volume, as was done in \citealp{Driver.etal.2005}. However, as we discuss in Section \ref{sec:characterising}, the analysis we perform here does not require this.

\section{Characterising the Local Volume}
\label{sec:characterising}

\begin{figure*}
   \centering
   \includegraphics[width=0.98\textwidth]{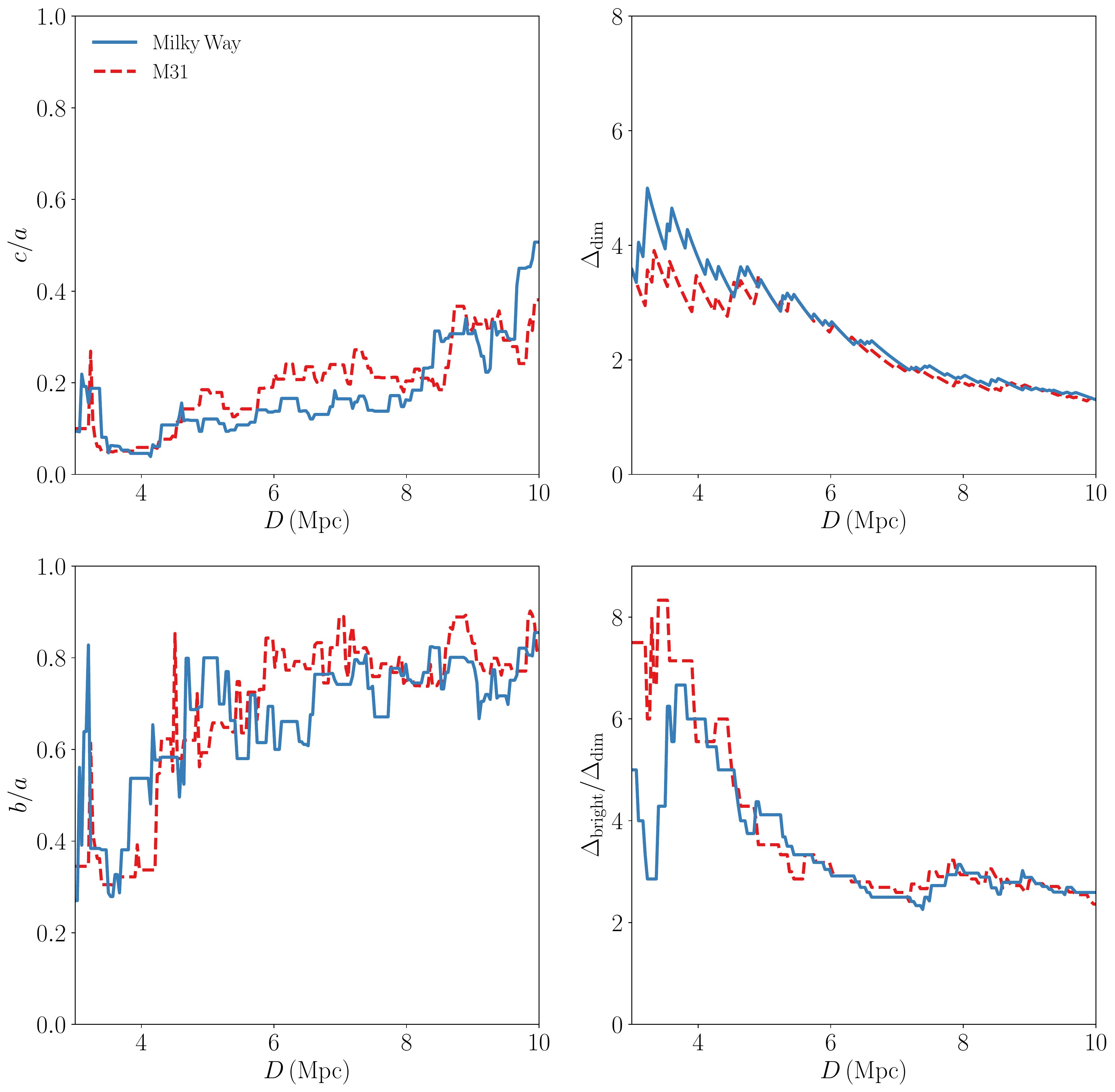}
\caption{Properties of the Local Volume as a function of distance from the Milky Way and M31. Top left: the ellipsoidal axis ratio $c/a$ of the distribution of galaxies. Top right: the overdensity $\Delta$ of dim ($-18<M^0_B<-16$) galaxies. Bottom left: the ellipsoidal axis ratio $b/a$ of the distribution of galaxies. Bottom right: the ratio $\Delta_{\rm bright} / \Delta_{\rm dim}$, which shows the overabundance of bright galaxies relative to dim galaxies.}
\label{fig:LV_fcn_D}
\end{figure*}

\begin{figure}
    \centering
    \includegraphics[width=0.47\textwidth]{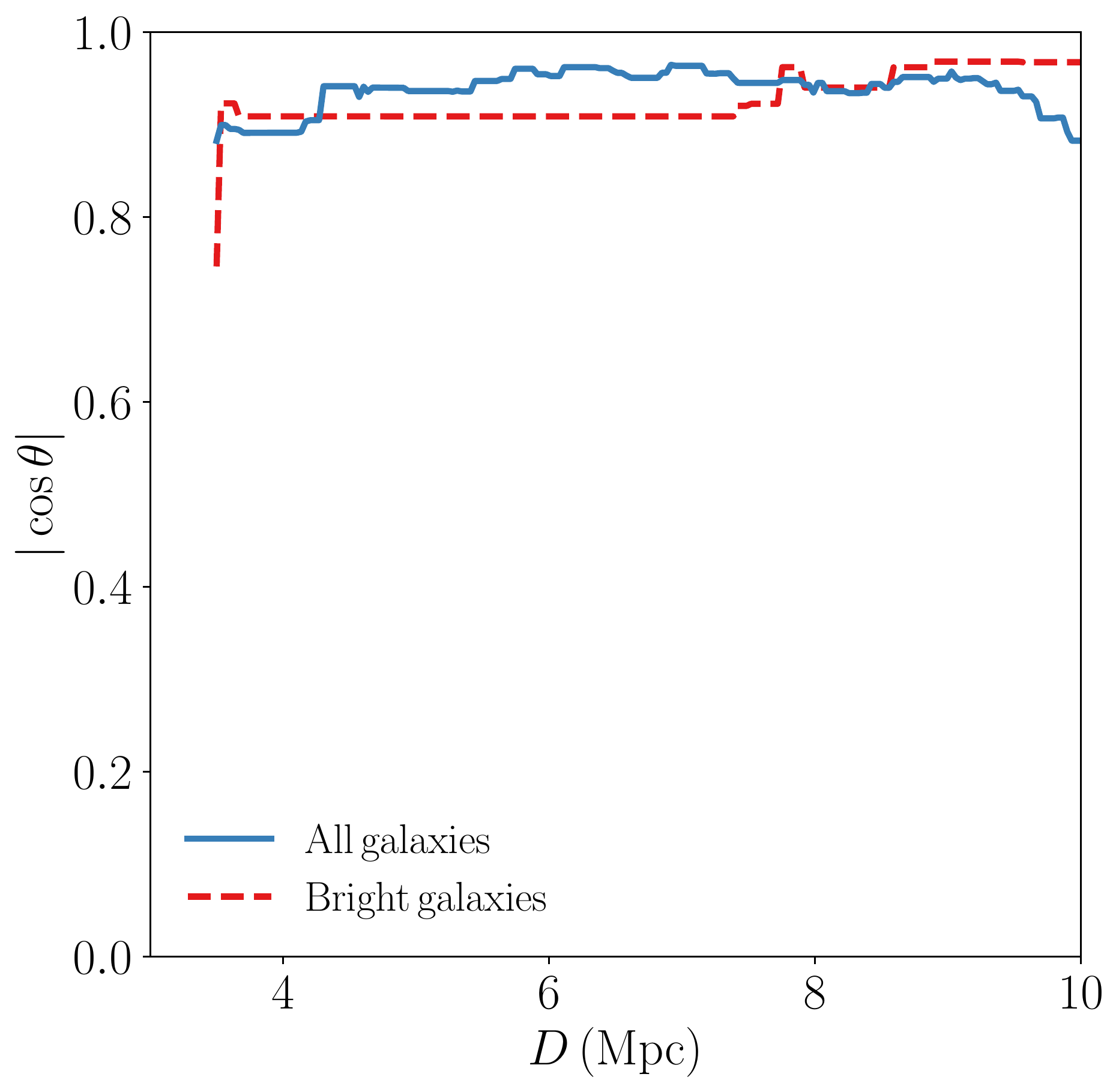}
    \caption{\color{black}The cosine of the angle $\theta$ between the short axis of the flattened structure of the Local Volume measured within distance $D$ and the short axis of the large-scale shear tensor of the Local Supercluster estimated by \citet{Libeskind.etal.2015}.  The blue curve uses the short axis measured for all Local Volume galaxies  brighter than $M_B^0=-16$ and a surface brightness $\leq 24.2$, and the red curve uses short axis estimated using `bright' galaxies as defined in Section \ref{sec:characterising}.}
    \label{fig:evec}
\end{figure}

As discussed in Section \ref{sec:intro}, some of the most readily apparent features of the galaxy distribution in the Local Volume are its flat sheet-like shape and its large concentration of bright galaxies. Although there are many potential characteristics of the galaxy distribution in the Local Volume one could examine, to avoid the risk of overspecifying we choose to focus on the simplest properties that capture these salient features.
\begin{table}
  \centering
  \caption{Properties of the Local Volume within 8 Mpc of the Milky Way and M31. The quantities $c/a$ and $b/a$ are the axis ratios of the ellipses, and $\Delta_{\rm dim}$ and $\Delta_{\rm bright}$ are the density contrasts of dim and bright galaxies relative to the MGC spectroscopic survey, respectively. These quantities are centred on both the Milky Way and M31 to give a simple estimate of the robustness of these measurements to small displacements.}
  \label{tab:LV_props}
  \begin{tabular}{c c c c c c}
  \hline\hline
  Galaxy & $c/a$ & $b/a$ & $\Delta_{\rm dim}$ & $\Delta_{\rm bright}$ \\ 
  \hline
  Milky Way & 0.163 & 0.786 & 1.699 & 5.190 \\
  M31 & 0.204 & 0.751 & 1.605 & 4.718 \\
  \hline
  \end{tabular}
\end{table}

To characterise the density of the Local Volume, we measured the density contrast relative to the MGC survey within spheres of radius $R.$ However, the density of galaxies in the Local Volume is strongly dependent on the magnitude limit used. So, we investigate the density contrasts $\Delta_{\rm bright}$ and $\Delta_{\rm dim}$, for bright galaxies having absolute magnitudes of  $M^0_B<-20.5$ and faint galaxies with $-16>M^0_B>-18$. Namely, we study the density contrast $\Delta = n/\bar{n}$, where $n$ is the number density of a particular galaxy sample and $\bar{n}$ is the mean number density of field galaxies within the same absolute magnitude and surface brightness ranges in the MGC sample. The `bright' magnitude limit was chosen to select galaxies with absolute magnitudes comparable to both that of the Milky Way, $M_B^0\approx -20.7$  \citep{Licquia.etal.2015,Bland_Hawthorn.2016}, and to the characteristic absolute magnitude of the Schechter form of the luminosity function, $M_\star = -20.34$ (see Section \ref{sec:mgc}). The magnitude range of the faint sample was chosen so that the Local Volume's density contrast is approximately constant throughout it. {\color{black} The impact of these choices are discussed in Section \ref{sec:unusual} (See Fig.~\ref{fig:ratio_fcn_MB})}.

{\color{black} Traditional methods for measuring the shapes of point distributions \citep[e.g.][]{Zemp.etal.2011} are inappropriate for the Local Volume. As we demonstrate in Appendix \ref{appendix:MDT}, simply measuring the eigenvalues of the unweighted shape tensor leads to measurements which are noisy and dominated by a small number of outliers. Commonly used methods for minimizing the impact of outliers fail catastrophically: introducing a $1/r^2$ weighting causes the measurement to be dominated entirely by the satellites of the MW, and iteratively recomputing the shape within ellipsoidal boundaries requires too many points. In Appendix \ref{appendix:MDT}, we introduce an alternative method for measuring axis ratios which mitigates these issues.} We note that our conclusions about the incidence of environments with LV-like properties in $\Lambda$CDM are not sensitive to the method chosen to measure axis ratios. We use all galaxies with $M_B^0 < -16$ to make shape measurements, the same lower limit as used by our dim sample. We use the standard notation for axis lengths with $a$, $b$, and $c$ referring to ellipsoid axis lengths in decreasing order. Thus, a highly flattened oblate ellipsoid has small $c/a$ and $b/a$ close to unity.

\begin{figure}
    \centering
    \includegraphics[width=0.47\textwidth]{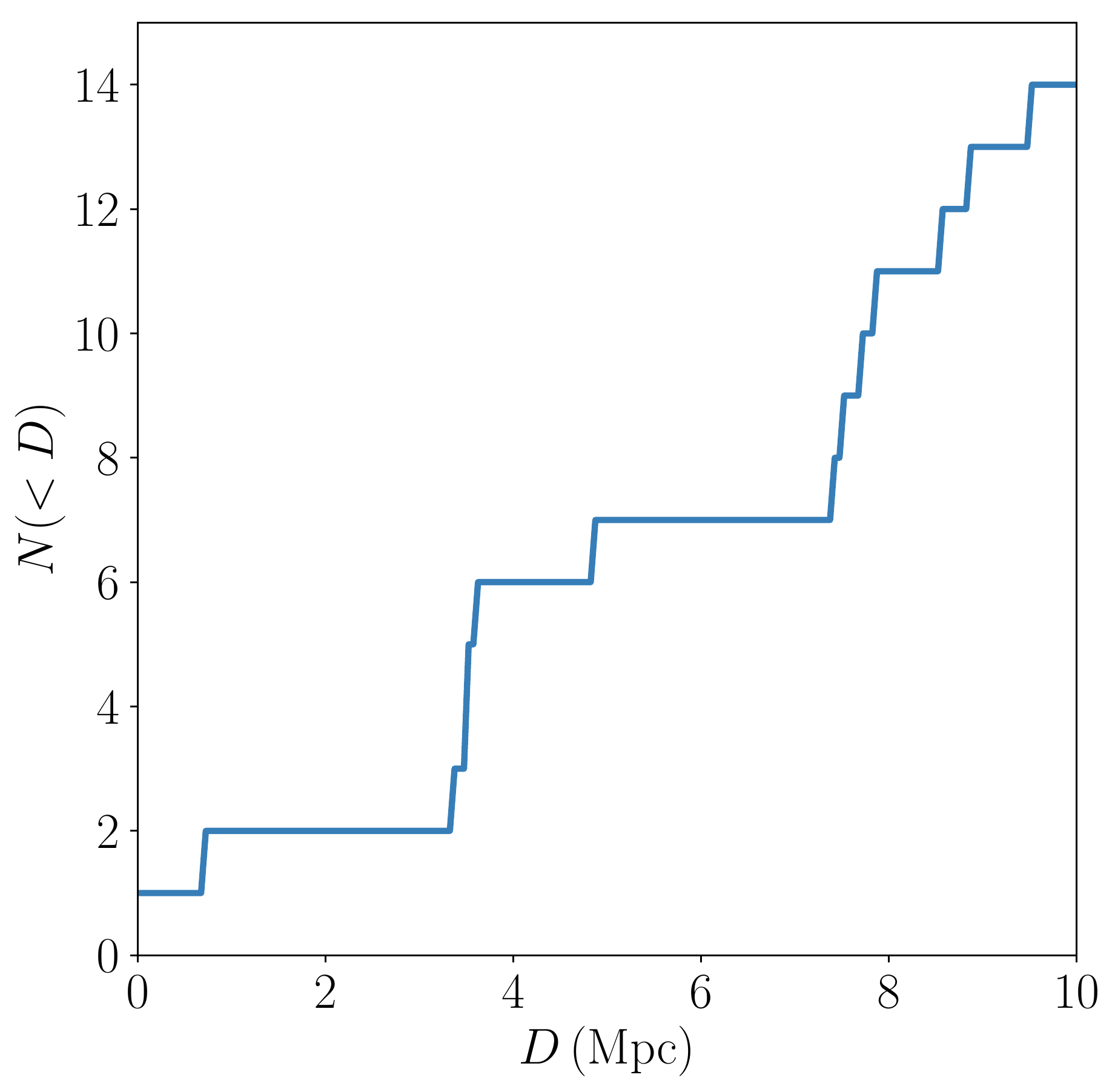}
    \caption{The cumulative distribution of bright galaxies ($M_B<-20.5$) in the Local Volume, out to 10 Mpc from the Milky Way. A concentration of galaxies is apparent between $3-5$ Mpc, which was referred to as the `Council of Giants' by \citet{McCall.2014}. Another concentration is present at $D>7.5$ Mpc.}
    \label{fig:brightgals}
\end{figure}

Fig.~\ref{fig:LV_fcn_D} shows ellipsoidal axis ratios, the density contrast of dim galaxies, and the ratio of density contrasts of dim and bright galaxies as a function of distance from the Milky Way. We also show these properties for M31-centric distances to indicate how robust the measurements are to small displacements. We choose the distance of 8 Mpc as our fiducial definition of the sample for several reasons. First, the peculiar velocity dispersion of galaxies in the observed `Local Sheet' is small within $\approx 7-8$ Mpc, but increases sharply beyond this distance \citep{Tully.etal.2008,Karachentsev.etal.2015,Anand.etal.2019b}. Second, as is shown in Fig.~\ref{fig:LV_fcn_D}, the $c/a$ axis ratio of the galaxy distribution also increases rapidly beyond this radius. Lastly, \citet{Mansfield.Kravtsov.2019} showed that density estimated in spheres of radius $R\approx 7 -8$ Mpc correlates with halo ages more strongly than density measured at any other scale.

{\color{black} We also compare the orientation of the sheet-like distribution of the Local Volume galaxies with the orientation of the eigenvectors of the large-scale shear tensor calculated in \citet{Libeskind.etal.2015}. Fig.~\ref{fig:evec} shows the angle between the short axis of the sheet, calculated within a given distance $D$ from the Milky Way, and the shortest eigenvector of the shear tensor. The figure shows that the short axis of LV galaxies is very well aligned with the short axis of the Local Supercluster. In Supergalactic coordinates, the normal to the sheet estimated using galaxies within $8$ Mpc ($M_B^0<-16$, $SB<24.2$) is estimated to be $(-0.168, -0.012, 0.986)$.}

We summarise the properties of the Local Volume under our fiducial definition in Table \ref{tab:LV_props}. In addition, 
Fig.~\ref{fig:brightgals} shows the cumulative distance distribution of galaxies in the bright sample. A similar distribution of bright galaxies around the Milky Way was presented by \citet{Karachentsev.Telikova.2018}, although for galaxies selected using stellar masses estimated from their 2MASS $K$-band magnitudes. 

Fig.~\ref{fig:brightgals} shows four bright galaxies at distances of $3-5$ Mpc. These galaxies are referred to as the `Council of Giants' by \citet{McCall.2014}, although in Fig.~\ref{fig:brightgals} the Council is diminished by the revision of the distances to the Maffei 1 and Maffei 2 galaxies to $5.73$ Mpc \citep{Anand.etal.2019}. In fact, both Maffei 1 and Maffei 2 are dimmer than our -20.5 magnitude limit for the bright sample. This is because Maffei 1 -- the central galaxy of the Maffei group -- is a late-type galaxy, which has a low mass-to-light ratio at optical wavelengths. The NIR luminosity of Maffei 1 implies that it has one of the largest stellar masses in the Local Volume, but it cannot be included while still using the MGC luminosity function as a reference. We also note that the Council of Giants described in \citet{McCall.2014} includes galaxies down to $M_B^0\approx -19.5$, considerably fainter than the fiducial limit of the bright sample we use.

The concentration of bright galaxies in the Local Volume is  also apparent or noted in the results of several recent studies \citep{Karachentsev.Kudrya.2014,Klypin.etal.2015,Kourkchi.Tully.2017}.
As Fig.~\ref{fig:brightgals} shows, in addition to the Council of Giants there is a another concentration of galaxies at $>7.5$ Mpc that contains about half of the bright sample. To the best of our knowledge, there are no previous estimates of how frequently these sorts of bright galaxy concentrations are expected in a $\Lambda$CDM cosmology. Simply estimating Poisson errors in high-luminosity bins is not sufficient to estimate the significance due to the existence of halo bias, which induces correlations in number density fluctuations across halo mass bins \citep[see, e.g.,][for detailed discussion]{Hu.Kravtsov.2003}.

\section{Are the Properties of the Local Volume Unusual?}
\label{sec:unusual}

Once we have quantified the properties of the Local Volume, we can compare them to the predictions of $\Lambda$CDM. To do this, we compare the axis ratios and density contrasts of the bright and dim galaxies estimated in the Local Volume with the distribution of the corresponding properties for Local Volume analogues in the SMDPL cosmological simulation.

\begin{figure}
	\centering
	\includegraphics[width=0.47\textwidth]{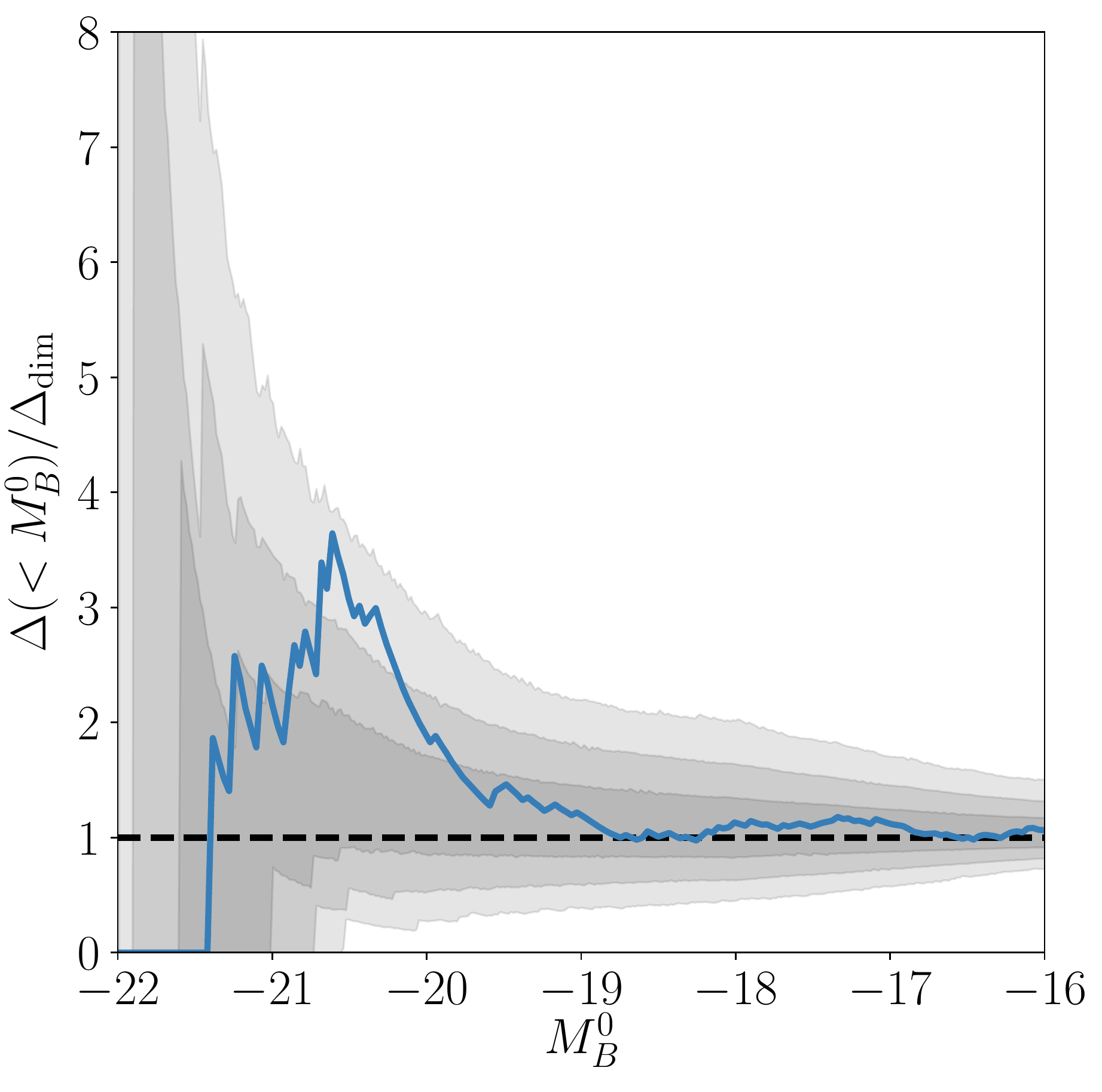}
    \caption{The density contrast of bright galaxies in the Local Volume relative to the density contrast of dim galaxies. The blue curve shows the density contrast brighter than the corresponding magnitude within $R=8$ Mpc of the Milky Way, normalised by the density contrast measured at $-18<M_{B,{\rm dim}}^0<-16$. The grey contours show the percentiles of the distribution corresponding to 1, 2, and 3 standard deviations for simulated Local Volume analogues. Here, Local Volume analogues are spherical volumes surrounding Milky Way-like haloes (as defined in Section \ref{sec:characterising}), which also have a Local Volume-like density contrast of faint galaxies ($1.5\leq\Delta_{\rm dim}\leq2$). Luminosity in the analogues is assigned by abundance matching against the MGC luminosity function.
    }
    \label{fig:ratio_fcn_MB}
\end{figure}

\begin{figure}
    \centering
    \includegraphics[width=0.47\textwidth]{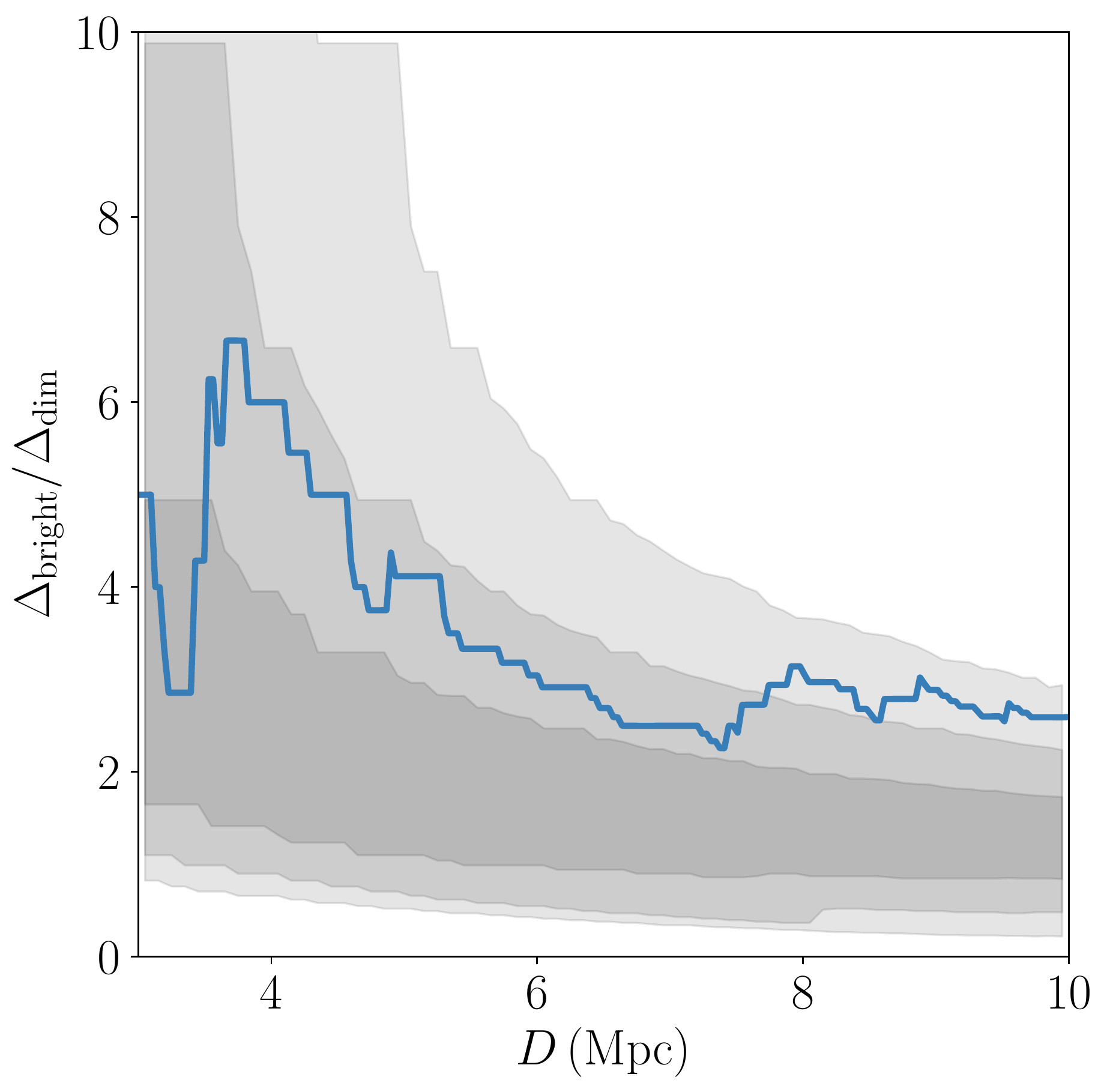}
    \caption{The same as Fig.~\ref{fig:ratio_fcn_MB}, but as a function of the outer distance used to define the Local Volume sample and LV analogues. The bright galaxy sample is defined using a constant $M_B^0 < -20.5$ cutoff at all distances. At and above our fiducial Local Volume definition, $D= 8$ Mpc, the $\Delta_{\rm bright}/\Delta_{\rm dim}$ fluctuation is at the $\approx2.5\sigma$ level. The bright galaxy fluctuation near the $\sim3-5$ Mpc `Council of Giants' identified by \citet{McCall.2014} is at the $\approx 1.5 \sigma$ level.}
    \label{fig:ratio_fcn_D}
\end{figure}

In SMDPL, Local Volume analogues are spheres centred on Milky Way analogues, which are defined to be {\color{black} isolated} haloes with absolute magnitudes assigned using abundance matching (see Section \ref{sec:abundance_matching}) in the range $-20.5>M^0_B>-21.25$ and  located $\geq 17$ Mpc from any halo with $M_{\rm vir}>1\times 10^{14} h^{-1}M_\odot$, {\color{black} where the mass threshold is chosen to be at the lower end of the range of virial mass estimates for the Virgo cluster \citep[e.g.,][]{Simionescu.etal.2017}.} The latter condition mimics the Milky Way's isolation from massive clusters. {\color{black} We do not require the presence of a cluster at a specific distance, but rather just require that no cluster be closer than the Virgo cluster is to the Milky Way.} We tested other typical requirements for Local Volume analogues, but deemed them to have little effect on the distribution of environment properties (see Appendix \ref{appendix:isolation_cuts}). {\color{black} After making these cuts, the sample contains 82971 Milky-Way analogues.}

The blue curve in Fig.~\ref{fig:ratio_fcn_MB} shows the density contrast of galaxies brighter than a given magnitude limit in the Local Volume relative to the density contrast of dim galaxies, $\Delta(<M_B^0)/\Delta_{\rm dim}$, using the dim magnitude range defined in Section \ref{sec:characterising}. This curve is compared against the distribution of equivalently defined ratios measured for the LV analogues in the SMDPL simulation with an additional requirement that $\Delta_{\rm dim}\in[1.5,2]$, similar to the $\Delta_{\rm dim}$ we estimated for the Local Volume.
The grey contours indicate the $1\sigma$, $2\sigma$, and $3\sigma$ levels of this distribution. 

Fig.~\ref{fig:ratio_fcn_MB} shows that there is a $\approx 2.5\sigma$ fluctuation of relative density contrasts at $M^0_B\approx -20.5.$ We show this fluctuation at our fiducial distance definition of $D=8$ Mpc in Fig.~\ref{fig:ratio_fcn_MB} and as a function of distance in Fig.~\ref{fig:ratio_fcn_D}. Both figures confirm that the Local Volume has an uncommon, but not exceedingly rare, overabundance of bright galaxies, which can be interpreted either as a fluctuation of bright galaxies or as a local change in the characteristic luminosity of the LF \citep[e.g.,][]{Eardley.etal.2015}. 
Fig.~\ref{fig:ratio_fcn_D} also shows that the $\sim3-5$ Mpc `Council of Giants' is a $\approx1.5\sigma$ fluctuation. It is the concentration of galaxies at $8-10$ Mpc that is considerably more unusual, making the Local Volume a $\approx 2.5\sigma$ outlier.

\begin{figure*}
   \centering
   \includegraphics[width=0.98\textwidth]{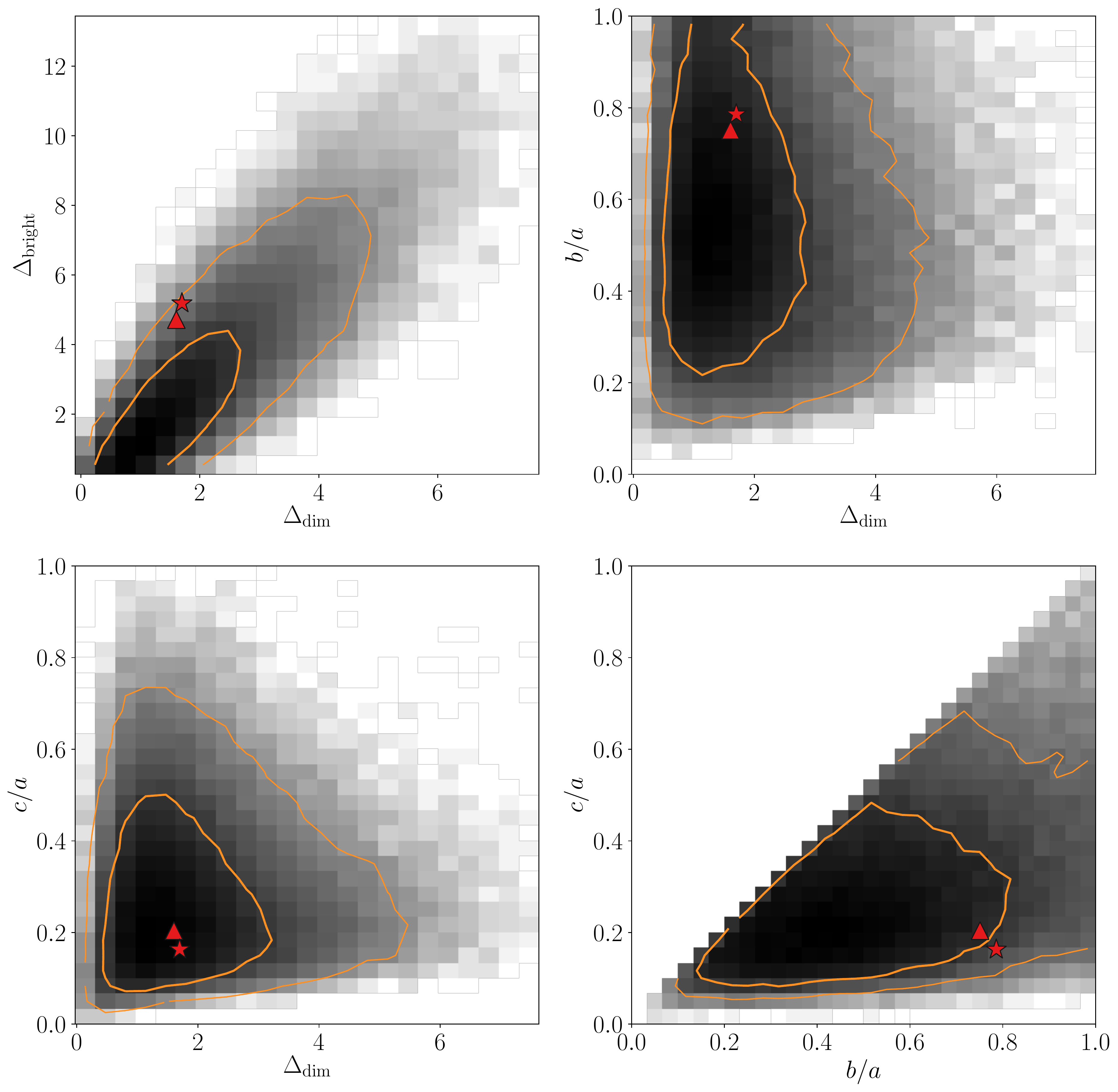}
   \caption{Distributions of properties of Local Volume analogues in the SMDPL simulation, abundance matched using the $M^0_B$ MGC luminosity function. Each spherical volume used to construct the distribution has radius 8 Mpc and is centred on a Milky Way analogue, which are taken to be haloes with $-21.25<M_B^0<-20.5$ that are farther than 17 Mpc from haloes with $M_{\rm vir}>1\times 10^{14} h^{-1}M_\odot$. Contours show the $95.45^{\rm th}$ and  $68.27^{\rm th}$ percentiles. Corresponding measurements for the Milky Way and M31 are indicated with red stars and red triangles, respectively. Top left: the density contrast of bright haloes, $\Delta_{\rm bright}$, versus that of faint haloes, $\Delta_{\rm dim}$. Top right: the axis ratio $b/a$ versus $\Delta_{\rm dim}$. Bottom left: the axis ratio $c/a$ versus $\Delta_{\rm dim}$. Bottom right: the axis ratio $c/a$ versus the ratio $b/a$.}
\label{fig:joint_distributions}
\end{figure*}

To investigate the likelihood of the Local Volume's other properties, we also show the 2D joint distributions of $\Delta_{\rm dim}$, $\Delta_{\rm bright}$, $c/a$, and $b/a$ for simulated LV analogues in Fig.~\ref{fig:joint_distributions}. These analogues are selected with the same criteria as above, but without restricting $\Delta_{\rm dim}$ to the range $[1.5,2]$ as was done in Figs.~\ref{fig:ratio_fcn_MB} and \ref{fig:ratio_fcn_D}. Fig.~\ref{fig:joint_distributions} shows that the Local Volume's $\Delta_{\rm dim}$ combined with each of its axis ratios is quite typical in $\Lambda$CDM. Both axis ratios together result in a significance of $\approx 1\sigma$, and the figure shows that the Local Volume is a $\approx 2\sigma$ outlier in the $\Delta_{\rm bright} - \Delta_{\rm dim}$ parameter space. 

To quantify the incidence of the LV-like environments, we present the fractions of LV analogues with probability densities smaller than that of the Local Volume in Table~\ref{tab:kde}. These numbers correspond to the probability isodensity contours that intersect the Local Volume values in the corresponding joint distributions.
To estimate probability density, we use ellipsoidal Epanechnikov kernels with axes scaled relative to the marginalised standard deviation, $s_i$, in each dimension, $i$. Our convergence testing showed that density estimates near the Local Volume had low sampling error and were independent of kernel axis length, $h_i$, for $h_i \sim 0.25\,s_i$ for our 2D distributions and $h_i \sim 0.75\,s_i$ for our 3D distributions, so we adopted these kernel sizes. Although all the probabilities cited in Table~\ref{tab:kde} use these kernel density estimates, we find that the 68.27\% and 95.45\% contours of our 2D distributions are unchanged if density is estimated by linearly interpolating 2D histograms.

The numbers in Table~\ref{tab:kde} show that the Local Volume is $\approx 2-2.5\sigma$ outlier in the distributions of $b/a-\Delta_{\rm bright}/\Delta_{\rm dim}$ and in $c/a-b/a-\Delta_{\rm bright}/\Delta_{\rm dim}$. Given the results in Figs.~\ref{fig:ratio_fcn_MB} and \ref{fig:ratio_fcn_D}, the main quantity determining these results is the relative density contrast of bright to faint galaxies, $\Delta_{\rm bright}/\Delta_{\rm dim}$.

A related question is the probability for simulated Milky Way-like haloes to have environments with properties that are more extreme than those of the Local Volume, such as $p(c/a\leq 0.163,b/a\geq 0.786)$ (see Table~\ref{tab:LV_props}). This is a separate question because the points lying along the same isodensity contour of the joint distribution can have combinations of $c/a$ and $b/a$ completely different from those of the Local Volume. This probability can be estimated by integrating the joint pdf in the region of $c/a \leq 0.163$ and $b/a \geq 0.786$ and dividing it by the integral over the entire pdf for normalisation. We find that the probability for the $c/a$ and $b/a$ values to be in these ranges is $\approx 0.0094$. We also  find that the probability for simulated environments to have an overabundance of bright galaxies in the ranges $\Delta_{\rm bright} \geq \Delta_{\rm bright,MW}$ and $\Delta_{\rm dim} \leq \Delta_{\rm dim,MW}$ is $\approx 0.0009$. Environments which require both the axis ratios and the overabundance to be in these ranges have a probability of $\approx 1.2 \times 10^{-5}$. 

However, one must be cautious when estimating and interpreting such probabilities. As we discuss in Section \ref{sec:comparison}, even environments with common properties can appear rare with such methodology, because unlike probability estimates based on isodensity contours, estimates based on integrating over rectangular regions will always decrease with increasing number of dimensions.

\begin{table}
  \centering
  \caption{The fraction of LV analogues in SMDPL with probability densities smaller than the Local Volume for a variety of properties. See Section \ref{sec:unusual} for further discussion. The reported errors are estimated from bootstrap resampling and do not account for measurement uncertainties or modelling uncertainties from abundance matching.}
  \label{tab:kde}
  \begin{tabular}{ll}
  \hline\hline
    Variables & $p$ \\
  \hline
  $\Delta_{\rm bright}$, $\Delta_{\rm dim}$ & 0.078 $\pm$ 0.004 \\
  $\Delta_{\rm bright}/\Delta_{\rm dim}$, $\Delta_{\rm dim}$ & 0.036 $\pm$ 0.004 \\
  $c/a$, $\Delta_{\rm dim}$ & 0.83 $\pm$ 0.01 \\
  $b/a$, $\Delta_{\rm dim}$ & 0.55 $\pm$ 0.01 \\
  $b/a$, $\Delta_{\rm bright}/\Delta_{\rm dim}$ & 0.016 $\pm$ 0.001 \\
  $c/a$, $b/a$ & 0.20 $\pm$ 0.01 \\ 
  $c/a$, $b/a$, $\Delta_{\rm bright}$ & 0.082 $\pm$ 0.005 \\
  $c/a$, $b/a$, $\Delta_{\rm bright}/\Delta_{\rm dim}$ & 0.022 $\pm$ 0.002 \\

  \hline
  \end{tabular}
\end{table}

\section{Discussion}
\label{sec:discussion}

Here we discuss the results of our study in the context of previous work, the effect of biases in the $B$-band magnitudes we used, some immediate implications for nearby satellite searches, and the prospects for selecting Local Volume analogues in cosmological simulations.   

\subsection{Comparisons with previous work}
\label{sec:comparison}

Our estimate of the density contrast of faint  galaxies can be compared with the results of \citet{Klypin.etal.2015}, who concluded that the number density of galaxies with $M_{\rm B}>-20$ was close to the mean number density of galaxies in this luminosity range at $z\approx 0$ (see their fig. 4 and associated discussion in their section 2.2). This is rather different from our estimate of the density contrasts, with $\Delta_{\rm dim}\approx 1.7$ and $\Delta_{\rm bright}\approx5.2$ within 8 Mpc and 
$\Delta_{\rm dim}\approx 3.5$ and $\Delta_{\rm bright}\approx 14$ within 5 Mpc of the Milky Way. This difference can also be seen directly in Fig.~\ref{fig:lum_func}. Our estimate thus indicates that the Local Sheet is a rather dense structure, especially if we take into account that overdensities are estimated in spheres, while the distribution of galaxies (and presumably mass) in the Local Sheet is flattened. 

We compared the luminosity function (LF) of LV galaxies that we measured to the LF estimated by \citet[][]{Klypin.etal.2015} and find that the shapes of the two LFs are in good agreement, but our LF is shifted to brighter magnitudes by $\approx 0.2-0.3$ mag. This is likely due mostly to the internal extinction correction that we apply to the magnitudes, although part of the difference may arise from the updated values of the $B_{\rm T}$ magnitudes that we use. Note that we have applied internal extinction corrections for our local LF and the reference LF of the MGC galaxies (see Fig.~\ref{fig:internal_extinction} and Sections \ref{sec:extinction} and \ref{sec:mgc}) for a fair comparison. An additional significant difference in the estimate of the local density contrast is in the reference LF. We use the MGC catalogue with $B$-band photometry close to the Johnson $B$-band used for the LV galaxies. Thus, although we apply a correction to bring the magnitudes to the Johnson $B$-band, this correction is only $\approx 0.1$ mag on average (see Section \ref{sec:mgc}). \citet{Klypin.etal.2015}, on the other hand, compare to LFs measured using $g$-band SDSS \citep{Blanton.etal.2005} and $b_J$ 2dFRGS photometry \citep{Norberg.etal.2002} which require corrections that are larger by a factor of two. Additionally, we use the same surface brightness cut in both our local sample and our reference LF, while \citet{Klypin.etal.2015} assume that the local sample is complete at all surface brightnesses and use a references LF which models the distribution of luminosities beyond its surface brightness limit.

Although \citet{Klypin.etal.2015} note that the abundance of bright galaxies is larger than the mean, they did not quantify the difference or its significance.
As mentioned in Section \ref{sec:intro}, the `jump' in the abundance of  galaxies above the knee of the luminosity or mass function was also noted by \citet{Karachentsev.Kudrya.2014} and \citet{Kourkchi.Tully.2017}, while \citet{McCall.2014} discussed the `Council of Giants' -- several bright galaxies at distances of $3-5$ Mpc from the Milky Way. None of these studies, however, estimated the overdensity of bright galaxies relative to the mean density of galaxies of similar luminosities. 

\citet{Goh.etal.2019} have studied the incidence of LV-like environments in the Bolshoi-Planck simulation \citep{Klypin.et.al.2011}
and concluded that such environments are exceedingly rare in $\Lambda$CDM.
Their procedure for identifying LV analogues in simulations was to first identify `wall' structures in the dark matter distribution using the morphological structure classification algorithm  of \citet{Aragon.Calvo.etal.2010} and to then select LV analogues among identified `wall' regions of radius $\approx 3-5$ Mpc using an additional set of criteria (see their section 3.2). One of the criteria used was that the density within the wall was $0.8-1.2$ times the mean density of the universe. In addition, they required that each wall contains at least one halo in the the mass range $0.7-1.3\times 10^{12}\, M_\odot$ and a number of other criteria such as the existence of a pair of haloes, the  separation between them, etc. We believe that the conclusion of \citet{Goh.etal.2019} about the extreme rarity of the Local Sheet was largely due to the restrictive set of many criteria and the method used to estimate the incidence of LV-like environments in $\Lambda$CDM in such a high number of dimensions. This is an important 
point because similar approaches for estimating the incidence of systems with specific properties arise in other contexts, such as estimating the rarity of `satellite planes'  \citep[see, e.g.][]{Pawlowski.2018}.

The issue is that the probability of LV-like environments is defined as the fraction of haloes within some set of selection criteria or, equivalently, by integrating the pdf within a small, high-dimensional rectangular volume. This fraction will decrease rapidly as the number of criteria (again, the dimensionality of the parameter space) increases. Thus, if the number of criteria is large and they are restrictive the probability is guaranteed to be small. This is akin to identifying a person by a set of characteristics: the more characteristics we use, the more uniquely we identify the person in a very large population. We illustrate this at the end of Section \ref{sec:unusual}.

To be fair, \citet{Goh.etal.2019} do consider the incidence of Local Volume properties using smaller numbers of restrictions. In particular, they consider the rarity of `wall'-type environments without additional restrictive constraints and point out that only $\approx 25$ per cent of Milky Way-sized haloes reside in wall regions, which is similar to our estimate that 20 per cent of LV analogues lie outside the isodensity contour that intersects the Local Volume in $c/a-b/a$ space (see Table \ref{tab:kde}). However, \citet{Goh.etal.2019} then point out that only $12$ per cent of walls have density contrast in the range $[0.8-1.2]$, which makes the Local Volume fairly rare in their assessment. However, using such density restriction suffers from the issue we described above.    

We emphasize that our approach to estimating the incidence of the LV-like environments is qualitatively different from that of \citet{Goh.etal.2019}. We construct the full distribution of the corresponding environmental properties around Milky Way-sized haloes in simulation. These distributions are well sampled in the SMDPL simulation and allow us to robustly measure the isodensity contours enclosing a given fraction of the total integral of the distribution. We can then find the isodensity contour which intersects the location of the Local Volume in this parameter space to gauge how rare it is in the $\Lambda$CDM cosmology. As long as the distribution is sufficiently sampled to identify the isodensity contours accurately, this approach does not depend on the dimensionality of the distribution. 
Indeed, the incidence of Local Volume analogues that we estimate  in the distribution of $c/a-b/a-\Delta_{\rm bright}/\Delta_{\rm dim}$ or its marginalised distribution of $b/a-\Delta_{\rm bright}/\Delta_{\rm dim}$ is comparable (see Table \ref{tab:kde}). 

It is using this approach that we conclude that Local Volume is $\approx 2-2.5\sigma$ outlier compared to the expectation of $\Lambda$CDM cosmology.
The most unusual property of the Local Volume that is the main factor behind this conclusion is its bright-to-faint galaxy contrast, $\Delta_{\rm bright}/\Delta_{\rm dim}$.

\subsection{The Impact of Incompleteness and Biases in RC3 Magnitudes}
\label{sec:RC3_bias}

As discussed at length in Section \ref{sec:photometry}, optical photometry in the Local Volume has been collected piecemeal over many decades with a variety of different measurement techniques. 

One potential bias in our measurements is the completeness of the local census of dim galaxies. If the sample is significantly incomplete at $-16 > M_B^0 > -18$, this would increase the ratio $\Delta_{\rm bright}/\Delta_{\rm dim}$, making the Local Volume appear more significant. However, as we show in Fig.~\ref{fig:lum_func}, the faint-end slope in the Local Volume matches the MGC spectroscopic survey.

{\color{black} The disc of the Milky Way heavily obscures objects within $5^{\circ}-10^{\circ}$ of the galactic plane. The extinction maps used in this paper are less accurate in this region, called the ``zone of avoidance,'' and the probability of detecting galaxies -- particularly moderate surface brightness galaxies -- is reduced. However, the galaxies missing in the Zone of Avoidance should not have significant effect on our results because $\pm 10^{\circ}$ zone of avoidance covers only 25 per cent of the LV by volume. This is a conservative estimate because the flattened distribution of Local Volume galaxies has a minor axis which is nearly perpendicular to the minor axis of the disc of the Milky Way.  Our estimate shows that the fraction of obscured volume in such a planar structure would be approximately 11 per cent.}

As discussed above, the fact that the MGC (and all spectroscopic surveys) are incomplete below a cutoff surface brightness results in systematically underestimated magnitudes for $M^0_B \gtrsim -19$. However, as can be seen in Figs.~\ref{fig:lum_func} and ~\ref{fig:ratio_fcn_MB}, the density contrast of the Local Volume relative to the MGC is fairly constant at $M_B^0 \gtrsim -18$, meaning that any density contrast measured for sufficiently dim galaxies or sufficiently low-mass haloes will be similar regardless of the absolute magnitude limit. The same is true for the shape of the Local Volume. Fortunately, brighter galaxies are unaffected by surface brightness cutoffs, allowing us to centre these Local Volume analogues on true Milky Way analogues. 

The inhomogeneity of Local Volume photometry is another potential bias. 
Our sample is dominated by asymptotic magnitudes measured in the RC3 catalogue \citep{RC3}. The asymptotic magnitudes in RC3 are estimated using empirical morphology-dependent growth curves designed to minimize surface brightness residuals for each Hubble index, $T$ \citep{Buta.etal.1995}. However, these growth curves are less flexible than those typically used by modern surveys, and the fits performed in RC3 are often done using sparse and/or small-radius aperture magnitudes, making it possible that RC3 magnitudes are systematically biased.

Indeed, \citet{Prugniel.Heraudeau.1998} found that RC3 magnitudes were systematically dimmer by 0.06 mag than magnitudes measured with growth curves which were linear combinations of exponential and de Vaucouleurs profiles. Furthermore,  \citet{FrancoBalderas.etal.2004} found that RC3 magnitudes were 0.12 mag dimmer than the total magnitudes they measured, while \citet{Young.2004} found an apparent magnitude-dependent offset, with RC3 magnitudes being 0.03 mag brighter compared to their estimates at dim magnitudes and up to 0.5 mag dimmer at the brightest magnitudes. However, their bright galaxy sample is small, and no other study has found such a large offset. We perform a similar test, cross-matching the RC3 against the high quality photometry in \citet{McGaugh_Schombert.2014}. We find that RC3 magnitudes are 0.06 mag dimmer. While a direct comparison between MGC and RC3 would be ideal for this study, we could only identify four MGC galaxies with $B_T$ entries in RC3, meaning that no statistically meaningful comparison can be done.

From this literature review, we can conclude that RC3 $B_T$ magnitudes are likely to be $0.06-0.12$ mag dimmer than more modern total magnitude estimates. To test the impact of such a bias on our results, we redid our analysis after applying shifts to all our RC3-based $B_T$ magnitudes ranging from zero to 0.15 mag. We find that systematic biases $\gtrsim0.06$ mag start to increase the significance of the overabundance of bright galaxies. With no systematic bias, 99.0 per cent of SMDPL analogues have $\Delta_{\rm bright}/\Delta_{\rm dim}$ smaller than that of the Local Volume. This increases to 99.7 per cent if there is a 0.06 mag bias and 99.83 per cent if there is a 0.12 mag bias. When we implement the more extreme bias model argued for by \citet{Young.2004}, the significance reaches 99.93 per cent.

Due to the uncertainty in the exact amount of bias in RC3 $B_T$ magnitudes, as well as the unknown biases in other significant magnitude sources, like pre-RC3 surveys or RC3 $m_B$ magnitudes, we do not attempt to explicitly homogenise the zero points of our different magnitude sources. However, we note that the significance of the local overabundance of bright galaxies could substantially increase when this is done, and that such homogenisation  may be necessary for future studies. An all-sky survey with homogeneous photometry, such as the survey expected from the WFIRST mission, would be ideal to put characterisation of the LV properties on firmer footing. 

\subsection{Targets for Satellite Searches}

A number of studies carrying out deep satellite searches around Milky Way analogues in the Local Volume are currenly underway \citep{Smercian.etal.2018,Cronjevic.etal.2019,Bennet.etal.2019,Carlsten.etal.2019}. This should significantly increase the number of systems with well-resolved satellite populations which will allow for higher-precision measurements of satellite radial distributions and the variation in satellite luminosity function amplitudes \citep{Bennet.etal.2019}. The expanded satellite samples should also result in better characterisation of the properties and incidence of the `satellite planes'  \citep{Pawlowski.2018}.

The high abundance of bright galaxies discussed in Section \ref{sec:unusual} means that there are a large number of targets for such searches. Using the updates to the LVG from this study, in Table \ref{tab:analogues} we list the galaxies within 11 Mpc and with $M_B^0 < M_\star$ -- i.e., properties similar to those included in the literature review presented by \citet{Bennet.etal.2019}. This distance threshold is within the range of TRGB distance measurements and the luminosity threshold selects galaxies comparable to or brighter than the Milky Way. 

Among the 27 galaxies listed in Table \ref{tab:analogues}, six already have deep satellite surveys, and four are within the Milk Way's zone of exclusion. The remaining 17 are good candidates for future satellite searches, three of which have already had a large number of candidate satellites identified by \citet{Carlsten.etal.2019}.

\begin{table}
  \centering
  \caption{The 27 bright ($M_B^0<M_\star$) Milky Way analogues within 11 Mpc categorised by their suitability for future satellite searches. Galaxies are ordered by increasing distance. Category (1) contains galaxies with existing deep ($M_{V,{\rm lim}} \leq -10$) satellite surveys. Category (2) contains galaxies which have extensive satellite candidate searches from \citet{Carlsten.etal.2019}. Category (3) contains galaxies which are within 15$^\circ$ of the galactic disc and would make poor candidates for searches due to heavy extinction. Category (4) contains early-type galaxies. Category (5) contains the remaining late-type galaxies. The galaxies in groups (2), (4), and (5) are prime targets for future satellite searches.}
  \label{tab:analogues}
  \begin{tabular}{cl}
  \hline\hline
    Category & Galaxy names \\
    \hline
    (1) & Milky Way, M31, CenA, M81, M101, M96\\
    (2) & NGC4631, M51, M104 \\
    (3) & IC0342, NGC6946, Maffei1, UGCA127\\
    (4) & NGC3115 \\
    (5) & NGC4845, M83, M106, NGC2903, M63, M108,\\
    & NGC1291, NGC0891, NGC2683, NGC6744,\\
    & M74, NGC3115, M66\\
  \hline
  \hline
  \end{tabular}
\end{table}

\subsection{The Prospects of Selecting Milky Way-Like Environments}
\label{sec:selecting}

As discussed in Section \ref{sec:intro}, modelling the Milky Way and its neighbours often requires selecting analogues from simulations. Most often analogues are selected simply to have virial masses similar to that of the Milky Way. Some studies include additional criteria, such as membership in a Local Group-like pair and isolation from Virgo-like galaxy clusters. As spectroscopic surveys, such as SAGA \citep{Geha.etal.2017} and DESI \citep{DESI.2016}, begin to allow deep studies of galaxy-scale satellite systems outside the Local Volume, similar observational selections will need to be made for certain classes of science targets.

Our study is an important component of such attempts because it characterises key components of the local environment. However, {\em which} features of the environment have the meaningful impacts on which galaxy and halo properties is a field of active research and is not yet settled, meaning that our work on its own cannot be fully prescriptive. Below we discuss several considerations for the ongoing effort to better select environments similar to the Local Volume.

Galaxy clusters generally affect their nearby environments significantly, making criteria based on galaxy cluster locality particularly meaningful. The large tidal fields of these clusters can stunt the growth of even distant haloes and are known to impact the formation history and characteristics of nearby haloes \citep[see review in][]{Mansfield.Kravtsov.2019}. In Appendix \ref{appendix:isolation_cuts} we show how such cuts affect the distribution environment properties. Additionally, studies of local velocity fields have shown that the presence of the Virgo cluster may affect the dynamics of galaxies in the Local Volume even at $\approx$17 Mpc away \citep[see, e.g.,][]{Shaya.Tully.2013,Shaya.etal.2017}.

Another common selection criterion requires that Milky Way analogues are isolated from similarly massive galaxies
\citep[with the possible exception of an M31-like counterpart, e.g.,][]{GarrisonKimmel.etal.2014,Griffen.etal.2016,Goh.etal.2019}. However, using such isolation criteria will likely select {\em against} regions with a high density of bright galaxies. There is nothing wrong about studying galaxies in these environments, but when making comparisons one has to be aware that such environments are systematically different from the Local Volume. In particular, the density of bright galaxies at distances $\gtrsim 8$ Mpc is known to correlate quite strongly with halo properties. For example, \citet{Mansfield.Kravtsov.2019} showed that dark matter haloes in regions with $\Delta(R=7.5\,{\rm Mpc})\gtrsim 2.16$ have truncated accretion histories and smaller scale radii due to the strong tidal fields and gravitational heating in these environments. 

The shape of the distribution of galaxies in the Local Sheet and Local Supercluster may also impact the properties of galaxies in the Local Volume, as large-scale tidal anisotropy is a strong predictor of halo properties \citep{Paranjape.etal.2018,Ramakrishnan.etal.2019}, the anisotropies of satellite distributions \citep{Libeskind.etal.2015}, and the amplitude of the satellite luminosity function \citep{Guo.etal.2015}, although the extent to which this is uniquely caused by large-scale anisotropy is unclear \citep{Goh.etal.2019,Mansfield.Kravtsov.2019}. Given that the distribution of galaxies in the Local Volume is uncommonly flattened  (See Table \ref{tab:LV_props} and Fig.~\ref{fig:joint_distributions}), the  most conservative practice for selecting Milky Way analogues would be to match the axis ratios of the local galaxy distribution in addition to the local overdensity. 

Thus, for comparisons with the Milky Way and other galaxies in the Local Volume, our results suggest selecting environments that are similarly flattened and are as crowded in bright galaxies as the Local Volume, while also being isolated from galaxy-cluster mass haloes, although the exact impact of each of these criteria requires further study. Given that such environments are fairly uncommon and may be difficult to identify for zoom-in resimulations of boxes with relatively small size, it is worth exploring approaches that can produce Local Volume-like environments by modifying cosmological initial conditions appropriately \citep[][]{Roth.etal.2016,Pontzen.etal.2017,Rey.Pontzen.2018}. 

\section{Conclusions}
\label{sec:conclusions}

In this study we quantify the  ellipsoidal axis ratios and overdensity of the distribution of faint and bright galaxies in the `Local Sheet.' We use the Local Volume Galaxy catalogue of \citet{Karachentsev.etal.2013}, in which we updated distances and apparent $B$-band magnitudes for many galaxies (see Section \ref{sec:local}).
We then compare the estimated axis ratios and density contrasts with the distribution of these quantities expected for the $\Lambda$CDM cosmology in Section~\ref{sec:unusual} using halo catalogues from the SMDPL simulation (Section~\ref{sec:sim}) with luminosities assigned using abundance matching (Section~\ref{sec:abundance_matching}). 
Our results and conclusions are as follows:

\begin{itemize}
\item The estimated axis ratios are $c/a\approx 0.16$ and $b/a\approx 0.79$ within 8 Mpc, indicating that the distribution of galaxies in the Local Volume can be approximated by a flattened oblate ellipsoid, consistent with the `sheet'-like configuration noted in previous studies \citep{Tully.etal.2008,Karachentsev.etal.2015}. 
\item In contrast with previous studies, which estimated that the Local Sheet has a density close to average, we find that the number density of faint galaxies ($-18<M^0_{B}<-16$) within 8 Mpc of the Milky Way is $\approx1.7$ times denser than the mean number density of galaxies of the same luminosity, while the number density of bright galaxies ($M_B^0<-20.5$) is $\approx 5.2$ times larger than the mean. This confirms the large abundance of bright galaxies in the Local Volume which has been qualitatively noted by previous studies (see Section~\ref{sec:intro}), but we also quantify this concentration  relative to the expectations of $\Lambda$CDM. 
\item Comparison with simulations (Section \ref{sec:unusual}) shows that the axis ratios and overabundance of bright-to-dim galaxies, $\Delta_{\rm bright}/\Delta_{\rm dim}$, of the  Local Volume are not exceedingly rare in the $\Lambda$CDM model, but they are not typical either. The estimated $\Delta_{\rm bright}/\Delta_{\rm dim}\approx 3$ alone makes our neighbourhood a $\sim 2.5\sigma$ outlier in the distribution of corresponding values in $\Lambda$CDM (Figs.~\ref{fig:ratio_fcn_D} and \ref{fig:joint_distributions} and Table \ref{tab:kde}), and is the most unusual property of the Local Volume that we considered.
\item The overabundance of bright galaxies is most significant at larger radii. Although the density contrast of bright galaxies relative to dim galaxies is largest within $4$ Mpc of the Milky Way, such an overabundance is more common in $\Lambda$CDM.
\end{itemize}

Our results indicate that the cosmic neighbourhood of the Milky Way is considerably more crowded with bright galaxies than is typical for galaxies of similar luminosity. The impact of the peculiar properties of our neighbourhood on the properties of the Milky Way and other nearby galaxies is not yet understood and warrants further study. 

\section*{Acknowledgements}
{\color{black} We would like to thank the referee of this paper, Noam Libeskind, for a thoughtful and very constructive report which resulted in a number of improvements  and clarifications in the presentation of our results}. We are also grateful to Brent Tully, Stefan Gottl\"ober, Vasily Belokurov and participants of the IoA Streams group for useful discussions during preparation of this paper.
MN was supported by the NSF REU program via grant NSF-PHY 1757898 to the University of Chicago. 
AK was supported by the NSF grants AST-1714658 and AST-1911111. PM was supported by the Kavli Institute for Cosmological Physics at the University of Chicago through grant PHY-1125897 and an endowment from the Kavli Foundation and its founder, Fred Kavli. 
The CosmoSim database used in this paper is a service by the Leibniz-Institute for Astrophysics Potsdam (AIP).
The MultiDark database was developed in cooperation with the Spanish MultiDark Consolider Project CSD2009-00064.
The MultiDark simulation project was supported by the Gauss Centre for Supercomputing e.V. (\url{www.gauss-centre.eu}) and the Partnership for Advanced Supercomputing in Europe (PRACE, \url{www.prace-ri.eu}), and the GCS Supercomputer SuperMUC at Leibniz Supercomputing Centre (LRZ, \url{www.lrz.de}).

The Millennium Galaxy Catalogue consists of imaging data from the
Isaac Newton Telescope and spectroscopic data from the Anglo
Australian Telescope, the ANU 2.3m, the ESO New Technology Telescope,
the Telescopio Nazionale Galileo and the Gemini North Telescope. The
survey has been supported through grants from the Particle Physics and
Astronomy Research Council (UK) and the Australian Research Council
(AUS). These data and data products are publicly available from
\url{http://www.eso.org/~jliske/mgc/} or on request from J. Liske or
S.P. Driver. 

Analyses presented in this paper were greatly aided by the free {\tt python} programming language and the following free software packages: {\tt NumPy} \citep{NumPy}, {\tt SciPy} \citep{scipy}, {\tt Matplotlib} \citep{matplotlib}, and \href{https://github.com/}{\tt GitHub}. We have also used the Astrophysics Data Service (\href{http://adsabs.harvard.edu/abstract_service.html}{\tt ADS}) and \href{https://arxiv.org}{\tt arXiv} preprint repository extensively during this project and the writing of the paper. We also acknowledge the usage of the HyperLeda database (http://leda.univ-lyon1.fr), the NASA/IPAC Extragalactic Database, which is funded by the National Aeronautics and Space Administration and operated by the California Institute of Technology, and the SIMBAD database,
operated at CDS, Strasbourg, France.



\bibliographystyle{mnras}
\bibliography{refs}



\appendix

\section{Estimating axis ratios}
\label{appendix:MDT}
\begin{figure}
   \centering
   \subfigure{\includegraphics[width=0.44\textwidth]{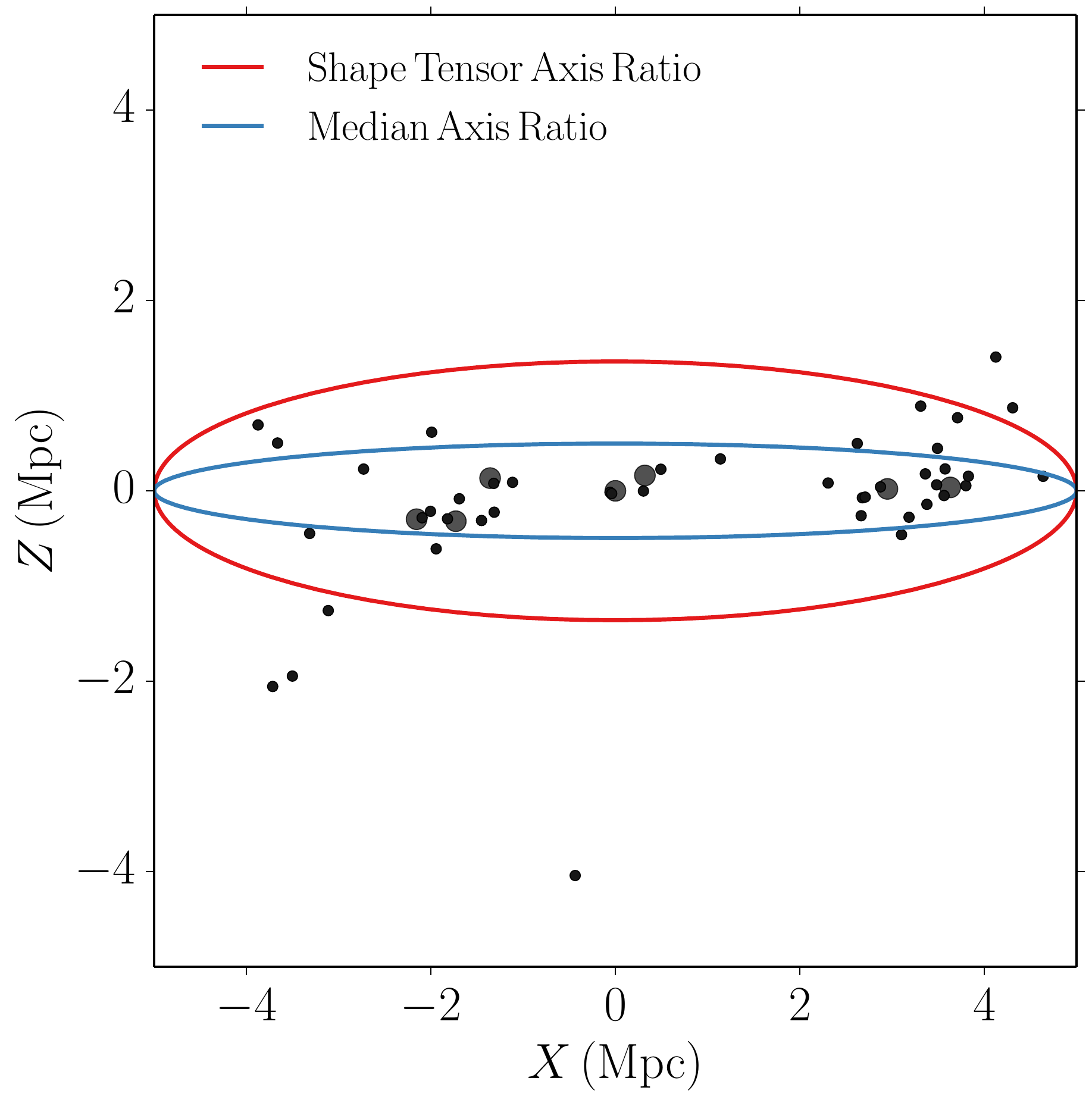}}
\caption{Comparison of two different methods of estimating ellipsoidal axis ratios for the distribution of galaxies within $5$ Mpc of the Milky Way. Large black circles show bright galaxies with $M_B^0 < -20.5$, and smaller black dots show dwarf galaxies with $M_B^0 < -16.$ These galaxies have been transformed into a coordinate system aligned with the minor and major eigenvectors of the distribution. The red and blue ellipses have the axis ratios inferred by the standard shape tensor method and by the median {\color{black} distance} method, respectively (see Appendix \ref{appendix:MDT} for details). Both ellipses have been scaled to have the same major axis length to aid comparison. The unweighted shape tensor estimate is  more sensitive to outliers at the outksirts of the distribution compared to the median method and therefore results in a less flattened ellipsoid. Typical modifications to the unweighted shape tensor method make it less sensitive to outliers, but are not possible in the Local Volume for reasons we describe in Appendix \ref{appendix:MDT}.}
\label{fig:shape_comparison}
\end{figure}

\begin{figure}
   \centering
   \subfigure{\includegraphics[width=0.47\textwidth]{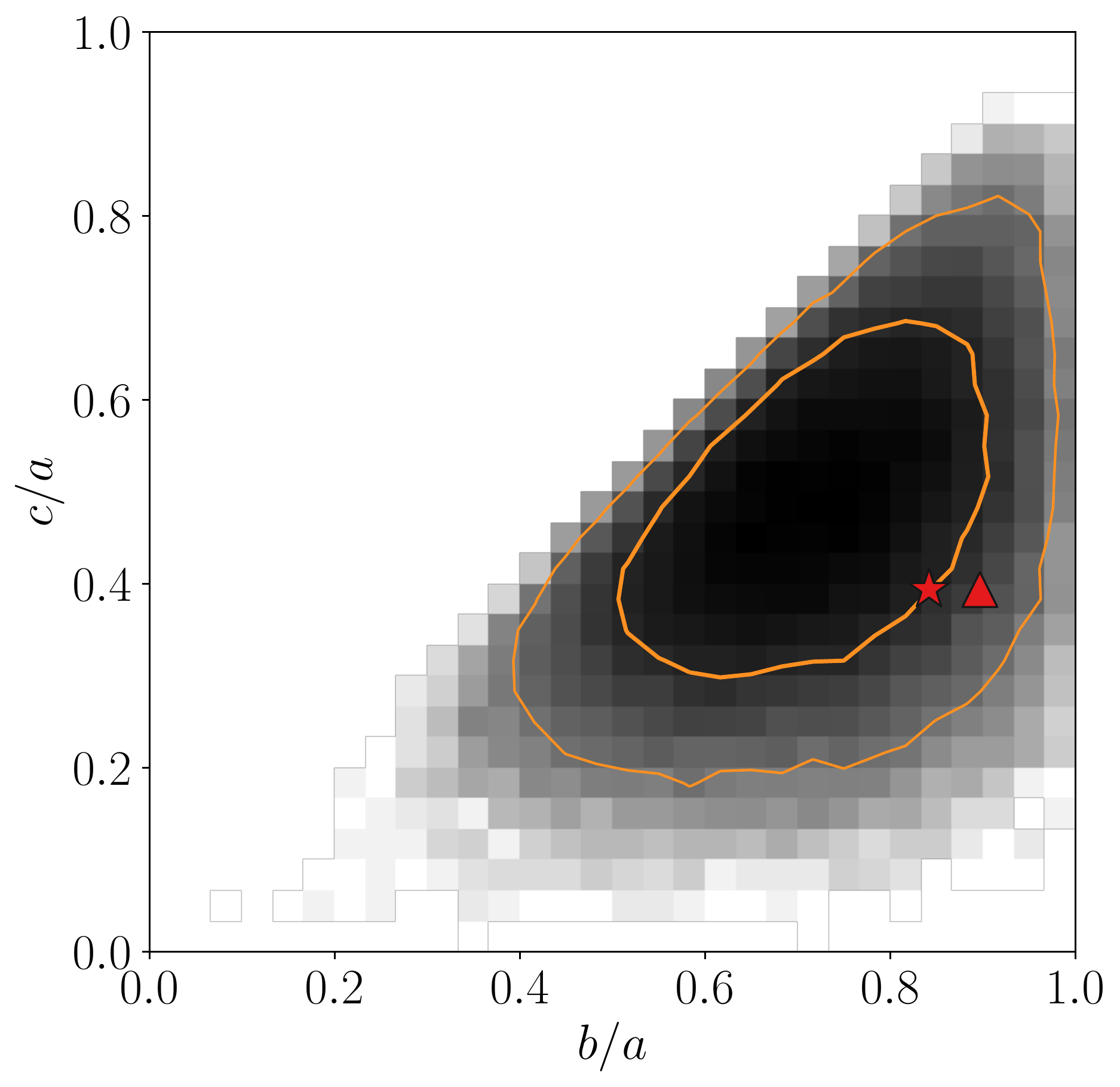}}
\caption{The same as the lower right panel of Fig.~\ref{fig:joint_distributions}, except using unweighted shape tensor eignevalues to measure axis ratios.}
\label{fig:shape_tensor_significance}
\end{figure}

The most common way to to estimate the shape of a distribution of point masses is to measure the eigenvalues of the shape tensor, $M_{ij}$\footnote{The shape tensor is also called the mass distribution tensor. The inertia tensor is defined differently and has different eigenvalues, but can be used in similar ways.}:
\begin{equation}
    M_{ij} = \sum_k ({\bf r}_k)_i({\bf r}_k)_jw_k.
\end{equation}
Here, $k$ indexes over points in a spherical volume, ${\bf r}_k$ is the displacement of the $k^{\rm th}$ point from the centre of the distribution and $w_k$ is the weight for that point, which we take to be unity.

In the limit of large $k$, a uniform-density ellipsoid with axis lengths $a_l$ will have the eigenvalues 
\begin{equation}
    \lambda_l=\left(\sum_k w_k\right) a_l^2/5.
\end{equation}

However, because each point is weighted by $r^2$, this method is very sensitive to outliers. We find that this sensitivity is particularly strong 
in estimating the shape of the galaxy distribution in the Local Volume. Fig.~\ref{fig:shape_comparison} shows that while this distribution is visually quite flattened, the axis ratio estimated from the unweighted shape tensor is less flat, primarily due to a small number of outlier galaxies.

There are two widely used methods for reducing sensitivity to objects at the outskirts of the distribution: assigning inverse-square weights, $w_k = r_k^{-2}$, and applying an iterative algorithm where subsequent shapes are measured using only points within the surface defined by the previous iteration (See \citealp{Zemp.etal.2011} for more details). Neither of these approaches is applicable to the Local Volume galaxies. Using $w_k=r_k^{-2}$ gives unphysical axis ratios as the shape tensor is completely dominated by the Milky Way's satellite distribution: when local axis ratios are measured as a function of limiting magnitude, there are sharp changes in the distribution every time a Milky Way satellite enters the sample. Even if the Milky Way's satellites are explicitly removed from the sample, the same becomes true for the satellites of M31. Additionally, we find that the iterative method does not consistently converge for point distributions as sparse as the ones considered here.

To estimate the shape of the galaxy distribution in the Local Volume we thus use  a different estimate for the axis ratios.  First, we measure the eigenvectors of $M_{ij}$ and transform our points into a coordinate system which aligns with these axes. We then measure the median distance from the origin along each axis $i$: $m_i = {\rm median}(|x_i|).$ We take the axis ratios of the distribution to be $m_i/m_j$ for each pair of axes $i,j.$ We favor this method for two reasons. First, it more accurately captures the qualitative shape of the Local Volume, as shown in Fig.~\ref{fig:shape_comparison}. Second, it is much less sensitive to outliers: the sampling error on axis ratios estimated by bootstrap resampling is a factor of 50 per cent smaller when using this median axis ratios method than when using a traditional method.

Although the median axis method is our preferred method for measuring axis ratios, this choice does not affect our significance estimates. In Fig.~\ref{fig:shape_tensor_significance} we compare the axis ratios of the Local Volume to the distribution of axis ratios measured in SMDPL using unweighted shape tensors. This is a direct analogue to the lower right hand panel of Fig.~\ref{fig:joint_distributions}. In both cases, the shape of the Local Volume is oblate and lies close to the $\approx1\sigma$ contour of the SMDPL distribution.

\section{The Effect of Isolation Criteria}
\label{appendix:isolation_cuts}

\begin{figure*}
   \centering
   \includegraphics[width=0.98\textwidth]{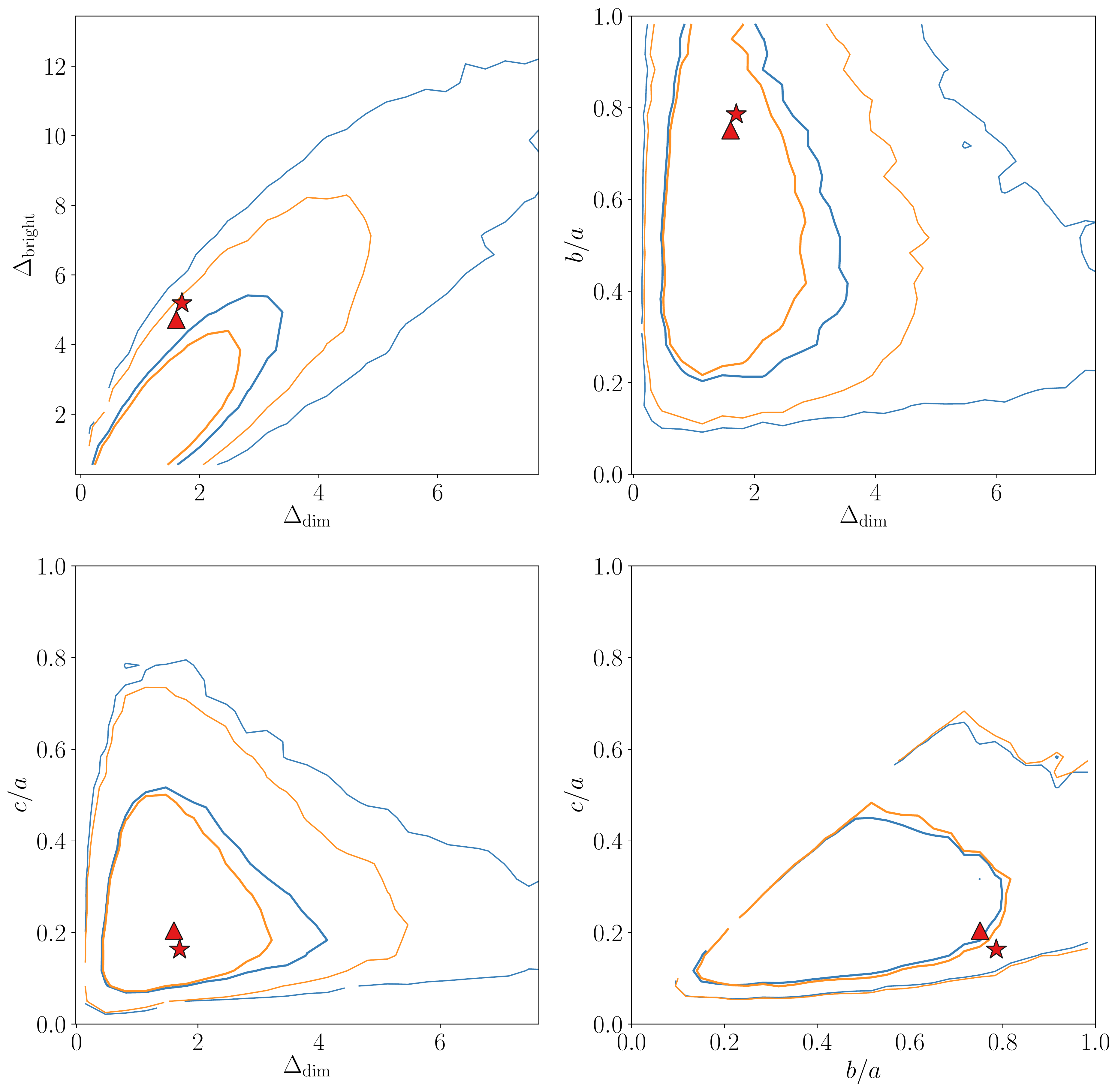}
\caption{The contours from Fig.~\ref{fig:joint_distributions} (orange), compared with the contours without any restrictions on clusters (blue). The stars marks the measurements for the Milky Way, and the triangles corresponds to M31. We also tested other restrictions: one, where the only requirement on the central halo was that it have $M^0_B<-20.5$ (without limiting the maximum brightness to -21.25 and without any restrictions based on proximity to clusters), and another, where the restrictions on the orange contour were combined with the requirement that the halo must have an M31-like pair ($-21.25<M^0_B<-20.5$) within 1 Mpc. The former resulted in curves similar to the blue, and the latter similar to the orange, though noisier due to number statistics.}
\label{fig:iso_criteria}
\end{figure*}

For our fiducial analysis, we identify Milky Way-like analogues in SMDPL simulation using  two criteria: the haloes must abundance match to $-20.5 > M^0_B > -21.25$, and they must be  $\geq 17$ Mpc away from haloes with $M_{\rm vir}>1\times 10^{14} h^{-1}M_\odot$. We ran our analysis with more lenient and more stringent conditions for these Milky Way analogues to determine whether specific choices affect our results.  Fig.~\ref{fig:iso_criteria} shows comparison of the properties of the Local Volume and environments of the Milky Way analogues in the SMDPL simulation without the cluster isolation requirement. The distributions of overdensities shift to larger vlaues, as expected given that objects in higher density environments around clusters are included, but there is no significant change in the conclusions about the incidence of the Local Volume-like environments.

We also tested the case when the only requirement was that the halo have $M^0_B<-20.5$; that is, without limiting the upper brightness to $-21.25$ and without cluster isolation criterion, and the results were similar to the blue curve in Fig.~\ref{fig:iso_criteria}. In addition, we combined our fiducial restrictions with the requirement that each Milky Way analogue have an M31-like neighbour ($-20.5<M^0_B<-21.25$) within 1 Mpc. This resulted in contours similar to the fiducial ones, although with larger noise due to smaller number of objects in the sample. Thus, these particular choices of isolation criteria for Milky Way analogues affect the resulting distribution of properties of the Local Volume counterparts, but do not have a significant effect on the conclusion about how common Local Volume-like environments are in the SMDPL simulation.  

\bsp	
\label{lastpage}
\end{document}